\documentclass[%
reprint,
superscriptaddress,
amsmath,amssymb,
aps,
pre,
floatfix,
longbibliography
]{revtex4-2}

\usepackage{times}
\usepackage{amstext}
\usepackage{graphicx}
\usepackage{color}
\usepackage{soul}
\usepackage{MnSymbol}
\usepackage{subfiles}
\usepackage{bm}
\usepackage{hyperref}
\usepackage{layouts}
\usepackage{enumitem}
\usepackage{natbib}
\usepackage{comment}
\usepackage{xcolor}
\usepackage{adjustbox}
\usepackage{float}
\usepackage{gensymb}
\usepackage[abs]{overpic}
\usepackage[normalem]{ulem}

\usepackage[T1]{fontenc}
\usepackage[utf8]{inputenc} 

\usepackage{etoolbox}
\usepackage{xparse}



\begin{document}

\title{Color Through Time: Distributions, Sparsity and Dominance in Western Painting}

\author{Ortensia Forni}
\email{ortensia.forni@polytechnique.edu}
\affiliation{{Econophysics Lab, Institut Louis Bachelier, 28 Pl. de la Bourse, Palais Brongniart, 75002 Paris, France}}
\affiliation{LadHyX UMR CNRS 7646, École Polytechnique, Institut Polytechnique de Paris, 91128 Palaiseau Cedex, France}

\author{Alexandre Darmon}
\affiliation{Art in Research, 33 rue Censier, 75005 Paris, France}
\affiliation{Capital Fund Management, 23 Rue de l’Université, 75007 Paris, France}

\author{Michael Benzaquen}
\affiliation{{Econophysics Lab, Institut Louis Bachelier, 28 Pl. de la Bourse, Palais Brongniart, 75002 Paris, France}}
\affiliation{Art in Research, 33 rue Censier, 75005 Paris, France}
\affiliation{Capital Fund Management, 23 Rue de l’Université, 75007 Paris, France}

\begin{abstract}
We investigate the evolution of color in Western painting using a large corpus of digitized artworks. Each painting is represented by its distribution over a fixed palette of colors in a perceptually uniform space. We compare the color distributions of historical periods and artistic movements through their Jensen--Shannon distances, and introduce two interpretable descriptors, sparsity and dominance, that quantify how many colors are used and how unevenly they are represented. We find a historical transition from dark, earthy palettes toward a broader and more saturated chromatic range, unfolding through distinct shifts from the nineteenth century onward. Periods evolve smoothly over time, whereas artistic movements display increasingly distinct chromatic signatures beyond chronology. In the sparsity--dominance plane, painting follows a two-stage historical trajectory, first broadening the range of colors used and later increasing color dominance. Comparison with natural images reveals distinct ways in which artistic palettes depart from nature, either through the colors they use or through how broadly and evenly those colors are distributed. Finally, two surveys on generated images show that judgments of harmonious color balance concentrate around two complementary regimes: many colors arranged in a strong hierarchy, or few colors used evenly. 
\end{abstract}

\date{\today}

\maketitle

\section{Introduction}
\label{sec:intro}

Color has always occupied a central place in artistic production, yet its study has long remained largely qualitative~\cite{gage1993colour}. From this perspective, color is not only a perceptual and material property, but also a carrier of symbolic, social, technical, aesthetic, and narrative functions. Its history has therefore been approached from very different angles, ranging from the vocabulary used to name colors to the chemistry of pigments and the social conventions governing their use~\cite{Histoire_des_couleur_Pastoureau}.
These perceptual and material dimensions also constrain any quantitative analysis. On the perceptual side, colors should be represented in a space that is approximately perceptually uniform, so that distances in that space reflect perceived differences. On the material side, the palette available to painters was historically constrained by the pigments and techniques accessible at a given time~\cite{philipp2003bright}.

The digitization of museum and archive collections has made large corpora of dated artworks directly measurable, and two broad strategies have emerged to describe them. The first relies on predefined, interpretable observables, such as the rank-ordered distribution of color usage and the fractal dimension of the palette~\cite{kim2014large}, permutation entropy and statistical complexity, which together track transitions between major historical periods~\cite{sigaki2018history}, or scalar measures of color contrast~\cite{lee2018heterogeneity}. Related quantities have also been applied to {\oe}uvres of single artists, from Pollock to Van Gogh and Rothko~\cite{taylor1999fractal,berezhnoy2007vangogh,papia2025rothko}, and are now assembled in toolboxes for empirical aesthetics~\cite{redies2025toolbox}. Their main advantage is interpretability, at the cost of reducing a work to a limited set of numerical descriptors.
The second strategy learns relevant features directly from the images. This includes classifiers of artistic movements and individual artists~\cite{icoglu2004classification,falomir2018qartlearn,honna2024ukiyoe}, convolutional networks that order paintings in time without access to their dates~\cite{elgammal2018shape}, and models that predict the emotions conveyed by an artwork~\cite{sartori2015affective,chen2024emotion}. Such flexibility comes with a corresponding loss of interpretability~\cite{elgammal2018shape}.
Both approaches have also been used to infer quantities external to the artwork itself, including wellbeing~\cite{gorin2025stateoftheart}, auction prices~\cite{stepanova2019palettes,borisov2024pricing}, and even the nineteenth-century rise of air pollution, inferred from the declining contrast of object contours in paintings by Turner and Monet~\cite{albright2023pollution}. In these studies, however, color primarily serves as an input to an inference problem rather than as the object of study in its own right.

The color statistics of paintings have also been compared with those of natural scenes, through amplitude spectra and sparseness~\cite{Graham2007StatisticalRO}, hyperspectral measurements of color gamuts~\cite{montagner2016statistics}, and analyses of specific movements such as graffiti~\cite{feitosa2020graffiti}. A related line of work connects chromatic composition to visual preference. Observers prefer compositions that remain within about seven degrees of the original color gamut of a painting~\cite{nascimento2017colors}, an effect that survives spatial scrambling~\cite{albers2020independent}, and preference closely tracks perceived naturalness~\cite{nascimento2021naturalness}, although the color statistics associated with preference differ from those of natural scenes~\cite{nakauchi2022regularity}. In our previous work, we found that preferred separations between simple hue pairs align with hue distributions in natural landscapes~\cite{forni2026harmonious}. Similarly, as some of us showed in~\cite{Lakhal_2020}, the preferred structural complexity of black-and-white images correlates with that of natural images. Natural scenes therefore provide a reference both for the chromatic statistics of paintings and for aspects of their aesthetic appeal. How the distance between painting and nature has evolved over time, however, remains largely unexplored quantitatively.

In this work, we take the distribution of colors itself as the object of study. After introducing the corpus and the color representation (Sec.~\ref{sec:data}), we first examine how the usage of individual colors evolves across eleven periods of equal size (Sec.~\ref{sec:color_usage_art_history}). We then move from individual colors to complete palettes by representing each period and artistic movement as a probability distribution over a fixed palette in the perceptually uniform OKLab space~\cite{ottosson2020perceptual}. This representation preserves the full structure of color usage, rather than reducing it to a frequency ranking~\cite{kim2014large} or a single contrast index~\cite{lee2018heterogeneity}. We compare these distributions across periods and twenty-six artistic movements using Jensen--Shannon distances, allowing us to quantify how chromatic practices differ across time and styles (Sec.~\ref{sec:colordistances}).
Distances between distributions, however, indicate how different two palettes are without revealing the nature of that difference. To obtain a more interpretable description, we  introduce two complementary observables, in the spirit of earlier work reducing image structure to entropy- and complexity-based descriptors~\cite{sigaki2018history}. \textit{Sparsity} $s$ quantifies the fraction of the available palette that remains unused, whereas \textit{dominance} $d$, defined as the Gini coefficient of the active-color distribution, measures the extent to which color usage is concentrated on a small subset of hues. Together, these variables distinguish paintings with restricted palettes from those using many colors, and balanced palettes from those dominated by a few colors. We use the resulting $(s,d)$ plane to trace the historical trajectory of painting (Sec.~\ref{sec:sd}) and to compare periods and movements with natural scenes (Sec.~\ref{sec:nature}).
Finally, we ask whether these same descriptors are related to aesthetic preference. Following our previous experimental work on color harmony and structural complexity~\cite{forni2026harmonious,Lakhal_2020}, we conduct two surveys using images generated at controlled values of sparsity and dominance (Sec.~\ref{sec:survey}). 

\section{Data and Methods}
\label{sec:data}

\begin{figure}[b]
    \centering
    \includegraphics[width=0.85\linewidth]{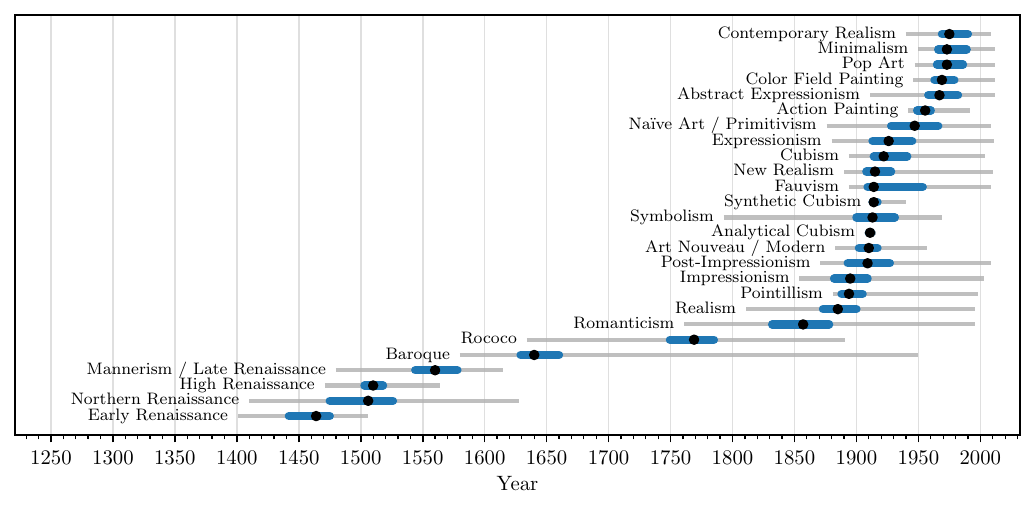}
    \caption{Chronological extent of the 26 artistic movements, computed from the 39,054 dated paintings. For each movement, the black dot marks the median production year, the thick blue bar the interquartile range, and the thin gray line the full range of production dates. Movements are ordered by median production year.}
    \label{fig:movement_temporal_span}
\end{figure}

We base our analysis on the \textsc{WikiArt} collection~\cite{wikiart_kaggle}, a dataset of digitized artworks labeled by artistic movement and painter and, for most works, by production year. Since our focus is on color, we first restrict the corpus to actual paintings. Alongside paintings, \textsc{WikiArt} contains a substantial number of drawings and sketches that are essentially monochrome and would therefore bias the color distributions toward neutral tones. We remove these automatically using a convolutional neural network fine-tuned to distinguish paintings from sketches and then applied to the full collection~\footnote{A ResNet-50 pretrained on ImageNet, with its final layer replaced by a two-class head; only that head and the last residual block were re-trained, on an $80/20$ train--validation split (cross-entropy loss, Adam optimizer with $\mathrm{lr}=10^{-4}$, batches of $32$, $10$ epochs). The classifier reached a validation accuracy of 96.56\%.}.
The remaining corpus is dominated by Western painting but also contains a sizeable group of Ukiyo-e works, whose distinct artistic tradition, materials, production techniques, and historical context make them less directly comparable within the same chronological framework. To retain a more coherent corpus and a consistent historical interpretation, we therefore restrict the analysis to Western painting and also exclude sculpture.
This leaves $69{,}486$ paintings spanning $26$ movements, from the Early Renaissance to Contemporary Realism; $39{,}054$ of them carry a reliable date, ranging from the fifteenth century to 2012. The chronological extent of each movement is shown in Fig.~\ref{fig:movement_temporal_span}. To compare these color distributions with those found in nature, we also use the Nature 12k landscape-photograph dataset~\cite{kaggle_12000} ($12{,}000$ images), processed through the same pipeline.

All color computations are carried out in the OKLab color space~\cite{ottosson2020perceptual}, an approximately perceptually uniform space in which Euclidean distance approximates perceived color difference; this is an essential property both for constructing the palette and for the nearest-color assignment below.
Rather than working with the full continuous gamut, we describe every image through a fixed reference palette of $n=25$ colors, shared across the entire study so that all distributions live in a common space. The choice of $25$ provides a natural compromise between chromatic resolution and perceptual distinguishability, while keeping the palette sufficiently coarse for the statistical quantities introduced below to remain well sampled. A substantially larger palette would describe each painting more finely, but would also reduce the perceptual separation between neighboring colors and increasingly fragment the empirical color distributions.
The palette is constructed to cover the visible gamut as uniformly as possible in perceptual terms, by selecting 25 well-separated colors from a regular grid of the sRGB cube (see Supplemental Information, Sec.~\ref{sec:paletteconstruction}). The resulting palette is displayed in Fig.~\ref{fig:color_distribution_time_windows} in the Supplemental Information.

For each painting, we draw a random sample of $N=10{,}000$ pixels, convert them to OKLab, and assign each pixel to its nearest palette color. Using a fixed sampling budget rather than all available pixels serves two purposes. The first is computational. The second, and more substantive, is to give every painting the same weight. The number of pixels in a digital reproduction reflects the resolution at which the work was photographed and, indirectly, the physical size of the original support. Since our unit of analysis is the artwork rather than the square centimeter of painted surface, a common sampling budget is therefore the natural choice.
The resulting histogram over the $25$ colors, normalized to unit sum, defines the image's \emph{color distribution}. Pooling the histograms of all works in a group, whether a movement or a time window, and renormalizing yields the corresponding group-level distribution used throughout the analysis. The pair $(n, N)$ constrains the range of values accessible to the two descriptors introduced in Sec.~\ref{sec:sd}; we shall discuss further this choice there once these quantities have been defined.

\section{Historical evolution of color usage}
\label{sec:color_usage_art_history}

Here we investigate how the chromatic content of Western painting has evolved over time. We partition the dated artworks into eleven chronological bins defined as quantiles of the year distribution, so that each contains a comparable number of paintings. Because dated works in our dataset are strongly concentrated in the nineteenth and twentieth centuries, this equal-count binning yields two broad early windows ($1401$--$1655$ and $1656$--$1861$) followed by progressively finer ones, down to roughly a decade for the most recent periods. This choice equalizes the statistical weight of each bin rather than its temporal length, ensuring that early and recent color distributions are estimated with comparable precision.

\begin{figure}[t]
    \centering
    \includegraphics[width=0.85\linewidth]{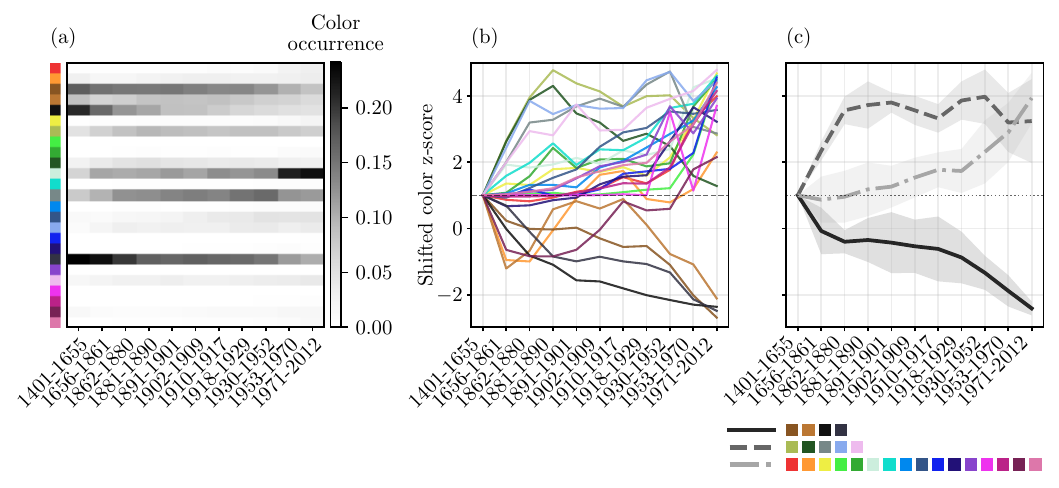}
    \caption{Evolution of palette color usage across the eleven equal-count periods.
(a)~Period-level color distributions $p_t(k)$. Each column is a period, each row one of the $n=25$ reference colors (swatches on the left), and the gray level gives the fraction of the period's sampled pixels assigned to that color, so that each column sums to one.
(b)~Shifted $z$-score trajectories (Eq.~\ref{eq:z_tilde}), one curve per reference color, each drawn in the color it represents; the dashed line marks the common starting value.
(c)~Mean trajectories of the three clusters obtained by Ward clustering of the curves in~(b): C0 (solid black; four dark earth colors), C1 (dashed dark gray; five muted mid-tones), and C2 (dash-dotted light gray; sixteen saturated hues and light tints). Shaded bands indicate $\pm1$ standard deviation across the member curves of each cluster. The palette colors belonging to each cluster are shown below the panel.}
    \label{fig:heatmap_zscore_clusters}
\end{figure}

Aggregating the palette counts of all paintings within a period and normalizing to unit sum yields a period-level color distribution $p_t(k)$, where $t=1,\dots,11$ indexes the periods and $k=0,\dots,24$ the palette colors. These distributions are displayed as a heatmap in Fig.~\ref{fig:heatmap_zscore_clusters}(a), where each column corresponds to a period and each row to a color of our palette.
The most visible feature is the dark, low-chroma, and earthy entries of the palette---black, a desaturated blue-black, browns, and ochres---carry most of the pictorial weight in the earliest periods and gradually fade thereafter. Beyond this broad redistribution, however, panel~(a) makes it difficult to compare the temporal evolution of colors that differ greatly in their overall abundance. In particular, changes affecting rare colors can remain visually muted.

To compare colors of very different overall abundance on the same footing, we compute for each color's period trajectory its $z$-score across the eleven periods:
\begin{equation}
    z_k(t) \;=\; \frac{p_t(k)-\mu_k}{\sigma_k},
    \qquad\text{where \ }
    \mu_k \;=\; \frac{1}{11}\sum_{t=1}^{11} p_t(k)
    \ \ \text{and } \ 
    \sigma_k^{2} \;=\; \frac{1}{11}\sum_{t=1}^{11}\bigl[p_t(k)-\mu_k\bigr]^{2}.
    \label{eq:zscore}
\end{equation}
This expresses every trajectory in units of that color's own historical standard deviation. We then shift all curves vertically to a common starting value,
\begin{equation}
    \tilde{z}_k(t)=z_k(t)-z_k(1)+1,
    \label{eq:z_tilde}
\end{equation}
as shown in Fig.~\ref{fig:heatmap_zscore_clusters}(b). The shift discards differences in initial level and retains only the \emph{shape} of the subsequent evolution, which is the feature we wish to compare across colors. A curve rising above $1$ thus marks a color that gains ground relative to its own occurrence in the earliest period, and a falling curve one that loses it. Note that the vertical axis carries no information about absolute abundance, which is read from panel~(a). The period-by-period distributions underlying this figure are reported in the Supplemental Information (Fig.~\ref{fig:color_distribution_time_windows}).
Because each curve in Fig.~\ref{fig:heatmap_zscore_clusters}(b) is rescaled by its own variability, the trajectories of even the rarest colors become clearly legible. This reveals patterns that are difficult to discern in panel~(a), including the late rise of several saturated hues.

While some colors decline steadily throughout, such as the dark earths (browns, ochres, and blacks), others rise early, before about $1880$, and then level off, including muted mid-tones such as grays, pale pink, green, and light blue. A third set, comprising the saturated hues (pure reds, greens, and blues), remains comparatively flat until the late nineteenth century and then rises through the twentieth. The trajectories therefore appear to organize into a small number of coherent temporal patterns, differing mainly in the timing of their changes. We make this structure explicit by clustering the $25$ colors on their shifted $z$-score curves.

Figure~\ref{fig:heatmap_zscore_clusters}(c) reveals three well-separated clusters of color trajectories, obtained by Ward clustering~\cite{ward1963,murtagh2014ward} of the shifted $z$-score curves in panel~(b). 
The first cluster (C0, solid black line) contains four dark, earthy colors---brown, ochre, carbon black, and a desaturated blue-black---whose relative frequency declines monotonically over time. The second (C1, dark gray dashed line) groups five intermediate, muted, or pastel tones: grays, a pale pink, light and dark green, and light blue. It follows the opposite pattern, gaining most of its ground before $1880$ and leveling off thereafter. The third cluster (C2, light gray dash-dotted line) contains sixteen colors, dominated by saturated, chromatically pure hues---pure red, yellow, greens, cyan, blues, violet, and magenta---together with a few high-value tints. Its mean trajectory remains close to its initial level until the late nineteenth century and then rises steadily through the twentieth. The last two clusters thus reach comparable relative levels, but on markedly different timescales.

These quantitative trends admit two complementary art-historical readings, one concerning the colors painters \emph{could} use and the other concerning what they \emph{chose} to paint and how they deployed color. The first is material. For much of the history of Western painting, the accessible palette was dominated by earth pigments (ochres, umbers, siennas), carbon- and bone-blacks, and a limited number of costly mineral colors~\cite{Histoire_des_couleur_Pastoureau, pastoureau2005petit}. Many of the vivid, high-chroma colors that our analysis finds emerging only in the later periods---including chrome and cadmium yellows, viridian and emerald greens, synthetic ultramarine and cobalt blues, and new synthetic magentas and violets---became available through the nineteenth- and twentieth-century development of industrial pigment chemistry~\cite{philipp2003bright}. The rise of cluster C2 therefore broadly tracks the expansion of the physically available color gamut.
The second reading concerns subject matter and pictorial function. Until the nineteenth century, the academic hierarchy of genres favored religious, mythological, historical, and portrait subjects, often set in interiors or darker settings and constructed through chiaroscuro and dark grounds---practices that naturally generate abundant blacks, browns, and ochres, largely independently of pigment availability~\cite{wolfflin2012principles}. The decline of cluster C0 may therefore reflect not only changes in available pigments, but also the waning of these subjects and techniques. At the same time, the rise of landscape and modern-life painting, together with \emph{plein-air} practice, shifted pictorial motifs toward daylight, sky, water, and foliage, favoring blues and greens and, more generally, a lightening of the picture surface as white grounds increasingly replaced dark ones. In the twentieth century, finally, color itself increasingly became an autonomous pictorial element rather than merely a means of depiction: the flat, saturated fields of Color Field painting and the artificial colors of Pop Art deploy pure hues for their own sake, a development echoed by the increase in color dominance that we quantify for the most recent periods (see Sec.~\ref{sec:sd} for more details).

It should be stressed that color statistics alone cannot causally disentangle these two mechanisms, since pigment availability and the evolution of subject matter are historically confounded and often mutually reinforcing---portable paint tubes, for instance, facilitated both the use of new pigments and \emph{plein-air} painting. Our analysis instead shows that the transformation of the palette unfolded in several stages, with distinct groups of colors changing at different times.


\section{Temporal drift and stylistic differentiation of color distributions}
\label{sec:colordistances}

We now move from the evolution of individual colors to the comparison of entire palette distributions, asking how similar two epochs, or two artistic movements, are in their overall use of color. For any two palette distributions $p$ and $q$, we quantify their dissimilarity using the Jensen--Shannon distance $D_{\mathrm{col}}(p,q)$, a symmetric and bounded metric well suited to comparing probability distributions~\cite{lin1991}. 

\begin{figure}[t]
\centering
\includegraphics[width=0.85\linewidth]{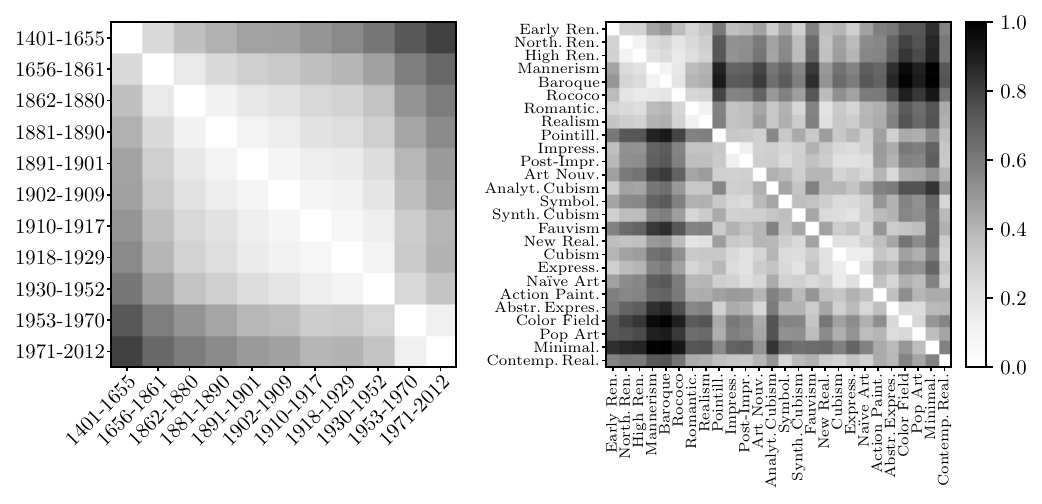}
\caption{Jensen--Shannon distances between color distributions. (left)~Pairwise distances between the eleven historical periods, ordered chronologically. (right)~Pairwise distances between the 26 artistic movements, ordered by median production year (Fig.~1).}
\label{fig:matrices_JS_colors_movs_years}
\end{figure}

Figure~\ref{fig:matrices_JS_colors_movs_years} shows the matrices of pairwise Jensen--Shannon distances between the eleven historical periods (left), ordered chronologically, and between the $26$ artistic movements (right), ordered by their median production year (see Fig.~\ref{fig:movement_temporal_span}).
These two matrices display strikingly different structures. The periods matrix is strongly banded, and the Jensen--Shannon distance grows almost monotonically with temporal separation, correlating well both with the number of periods separating two windows and with the elapsed time between them. The average palette therefore drifts gradually and cumulatively, with each period resembling its neighbors and departing progressively from more distant ones.
The movements matrix shows a very different organization. Rather than forming a smooth chronological gradient, the distances arrange into blocks, with chronology providing only a partial ordering. A pale block in the upper left groups the earliest schools, from the Early Renaissance to Realism, whose palettes remain remarkably similar despite spanning roughly four centuries and being sharply distinguished by art history in terms of composition, drawing, and iconography. This chromatic stability is consistent with the comparatively limited evolution of the pigments available over much of this period, which imposed a broadly shared material basis on otherwise very different artistic traditions.
At the opposite extreme, the lower left region, where the most recent movements are compared with the earliest ones, contains some of the largest distances in the matrix: twentieth-century palettes are among the furthest from those of the early block, consistent with the period matrix. Between these extremes, however, movements do not arrange along a simple temporal gradient. Chronologically adjacent movements can have very different chromatic distances, particularly among the more recent schools, where alternating lighter and darker rows reveal strong movement-specific differences that have no counterpart in the smoother period matrix.

We interpret this pattern as a possible change in the role of color. In the early block, color appears primarily functional to the depiction of the visible world: movements that differ markedly from one another draw on broadly similar palettes and are distinguished more by how they use the available pigments than by the colors themselves. From the beginning of the twentieth century, and particularly with the avant-garde movements of those years, color increasingly seems to become part of the signature of each movement. The industrial expansion of the palette, photography relieving painting of part of its role in recording appearances, and the growing independence of painters from commissions and codified subjects may all have contributed to this transformation. The distance matrices are consistent with these factors, without establishing a causal link.

Painters exploited this expanded freedom in very different directions. Analytical and Synthetic Cubism return to palettes close to those of the early centuries, visible as the lighter rows of the modern block, while radically renewing form and subject~\cite{golding1961cubism}. Fauvism, chronologically adjacent to them, lies among the darkest rows and renders recognizable scenes through deliberately altered color, making extensive use of the newly available pigments~\cite{elderfield1976wild}. These divergent choices help explain the alternation of light and dark rows observed among the recent movements.

Taken together, the two matrices suggest that color is not simply a function of epoch. Across the periods, palette composition drifts smoothly with time, whereas among movements this chronological trend is overlaid by strong stylistic differences~\footnote{Part of this contrast may reflect aggregation: each period pools works from several movements and therefore averages out some of their specificities.}. The spectral decomposition supports this interpretation (see Fig.~\ref{fig:matrix_colors_pca} in the Supplemental Information): for the periods, the first non-trivial mode closely recovers the chronological ordering, whereas for the movements it follows time only partially. Chronology therefore provides a strong organizing axis for period-level color distributions, but is insufficient to account for the diversity observed across artistic movements.

\section{The sparsity--dominance plane}
\label{sec:sd}

A distance between color distributions tells us how much two palettes differ, but not in what respect. Since a painting's color distribution is a 25-dimensional object---too rich to read or compare directly---we summarize it with two complementary and interpretable quantities that answer two distinct questions: \emph{how many} colors a work uses, and \emph{how evenly} it uses them. This low-dimensional representation is in the spirit of interpretable reductions of pictorial content such as the entropy--complexity plane of Sigaki et al.~\cite{sigaki2018history}.
The first quantity measures how much of the palette a painting leaves unused. Denoting by $N_{\mathrm{act}}$ the number of active colors, that is, colors carrying nonzero weight in the image's color distribution, we define the \emph{sparsity} as
\begin{equation}
s = 1-\frac{N_{\mathrm{act}}}{n},
\label{eq:sparsity}
\end{equation}
where $n=25$ is the palette size introduced above.
High sparsity thus indicates a restricted palette, whereas low sparsity indicates that a larger fraction of the available colors is used.

The second quantity, \emph{dominance}, measures how unevenly the active colors are used, irrespective of their number. It is defined as the Gini coefficient of the proportions $p_1\le\dots\le p_{N_{\mathrm{act}}}$ of the active colors, sorted in non-decreasing order and summing to one:
\begin{equation}
    d \;=\; 1 + \frac{1}{N_{\mathrm{act}}}
    \;-\;
    \frac{2}{N_{\mathrm{act}}}
    \sum_{i=1}^{N_{\mathrm{act}}}
    \left(N_{\mathrm{act}}-i+1\right)\,p_i .
    \label{eq:dominance}
\end{equation}
Here $d=0$ when the active colors are used in equal proportion, corresponding to a perfectly balanced palette, and $d$ increases as their usage becomes more uneven, approaching $1 - 1/N_{\rm act}$ when a single color carries almost all the weight. The same value of dominance can, however, arise from different palette distributions (see Fig.~\ref{fig:schematic_DvsA} in Supplemental Information). We therefore also computed the Lorenz asymmetry coefficient~\cite{damgaard2000describing}, which captures differences in the shape of the distribution that are not encoded by the Gini coefficient alone. In practice, however, the two quantities seem to be strongly correlated and display very similar trajectories across periods and movements (see Fig.~\ref{fig:DA} in Supplemental Information), indicating that the asymmetry coefficient adds little additional information here. We therefore restrict the analysis to $d$ for simplicity.

The palette size $n$ and the pixel budget $N$ of Sec.~\ref{sec:data} determine the region of the $(s,d)$ plane that can be reached. Sparsity takes values in steps of $1/n$, from $0$ to $1-1/n$. Dominance is bounded both by the number of active colors and by the finite number of sampled pixels, since every active color must contain at least one pixel. Its maximum is reached when $N_{\rm act}-1$ colors contain a single pixel each and the remaining color contains all the others, giving 
\begin{equation}
d_{\max}(s) \;=\; 1 - \frac{1}{N_{\rm act}} \;-\; \frac{N_{\rm act} - 1}{N}, \qquad\text{with \ } N_{\rm act}(s) = n\,(1 - s).
\label{eq:Dmax}
\end{equation}
The term $1-1/N_{\rm act}$ is the intrinsic ceiling of the Gini coefficient for $N_{\rm act}$ proportions and decreases as sparsity increases, from $0.96$ at $s=0$ to $0.5$ at $s=1-2/n$. A small palette would therefore both sample sparsity coarsely and leave a substantial part of the $(s,d)$ plane inaccessible. The term $(N_{\rm act}-1)/N$ is the correction due to the finite pixel budget. With $N=10{,}000$, it never exceeds $(n-1)/N=0.0024$, so the accessible region is practically indistinguishable from that obtained by using all pixels. Together with the perceptual considerations discussed in Sec.~\ref{sec:data}, this supports the choice $(n,N)=(25,10^4)$ as a reasonable compromise.

\begin{figure}[t]
    \centering
    \includegraphics[width=0.85\linewidth]{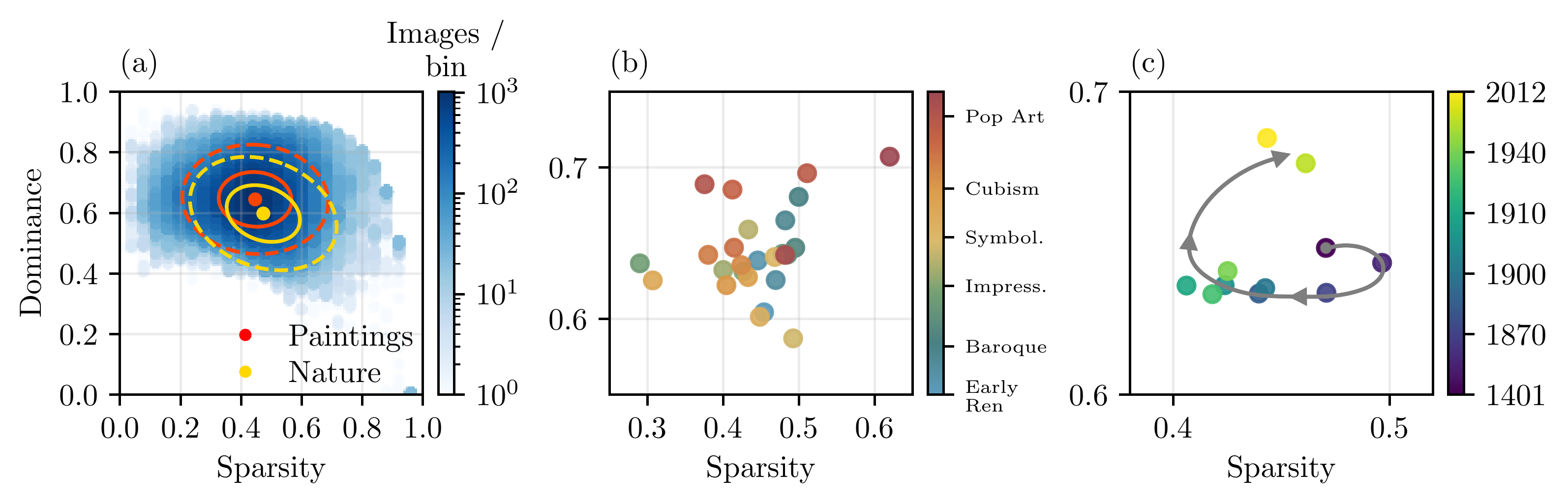}
    \caption{The sparsity--dominance plane. (a)~Sparsity and dominance of the 69,486 paintings; each point is colored by the number of paintings falling within its bin on a $50\times50$ grid (logarithmic scale). Solid and dashed ellipses indicate the one- and two-standard-deviation covariance ellipses of the paintings (red) and natural images (yellow), centered on their respective means, $(\bar s,\bar d)=(0.45,0.65)$ and $(0.47,0.60)$. (b)~Mean $(s,d)$ coordinates of the 26 movements, colored by rank in the median-year ordering of Fig.~1; selected movements are labeled on the color bar. (c)~Mean $(s,d)$ coordinates of the eleven periods, colored by chronological rank; the arrow indicates the direction of evolution over time, and the years on the color bar identify the corresponding periods.}
    \label{fig:SD_comparison}
\end{figure}

Mapping every painting onto the $(s,d)$ plane reveals that color usage is far from arbitrary (Fig.~\ref{fig:SD_comparison}). Figure~\ref{fig:SD_comparison}(a) shows that most of the $69{,}486$ artworks concentrate in a compact region---roughly $s\in[0.3,0.6]$ and $d\in[0.55,0.75]$---with a well-defined mode around the corpus averages $(\bar s,\bar d)=(0.45,0.65)$. Averaging the $(s,d)$ coordinates of the works within each artistic movement (Fig.~\ref{fig:SD_comparison}(b)) and within each historical period (Fig.~\ref{fig:SD_comparison}(c)) reveals the internal organization of this cloud. 
While the movement averages show no obvious ordering, the historical periods, by contrast, trace a clear and non-trivial trajectory in the $(s,d)$ plane, see the chronological arrow in Fig.~\ref{fig:SD_comparison}(c).
Such path has two successive phases. From the early peak in sparsity ($s\approx 0.50$ in $1656$--$1861$), the periods drift steadily \emph{leftward}, with sparsity falling to a minimum ($s \approx 0.41$ around $1910$--$1917$) while dominance remains essentially constant ($d \approx 0.63$--$0.64$). In other words, paintings therefore progressively recruit more of the available palette, a shift that may reflect the nineteenth-century industrialization of pigment production~\cite{philipp2003bright}, which widened the range of colors accessible to painters. In the second half of the twentieth century, the path then turns sharply \emph{upward}: dominance rises from $\approx 0.64$ to $0.68$ in $1953$--$1970$ and $0.69$ in $1971$--$2012$, while sparsity partly rebounds. This late rise in dominance is consistent with the growing importance of movements such as Pop Art and Color Field painting, whose compositions often rely on a small number of strongly dominant colors, and constitutes the two-dimensional counterpart of the surge of saturated hues documented in Sec.~\ref{sec:color_usage_art_history}.

\section{Distance from natural images}
\label{sec:nature}

A recurring question in the quantitative study of art and aesthetics is how the color statistics of paintings relate to those of natural scenes~\cite{Graham2007StatisticalRO,montagner2016statistics,nascimento2021naturalness,nakauchi2022regularity}. Since each painting is now characterized both by a color distribution (Sec.~\ref{sec:colordistances}) and by a point in the $(s,d)$ plane (Sec.~\ref{sec:sd}), we can make this comparison explicit in both representations. Using the natural-image dataset introduced in Sec.~\ref{sec:data}, we measure how far each artistic movement and historical period lies from natural images. The centroid of the natural images in the sparsity--dominance plane is $(s_{\mathrm{nat}},d_{\mathrm{nat}})=(0.47,0.60)$, compared with $(\bar s,\bar d)=(0.45,0.65)$ for paintings (Fig.~\ref{fig:SD_comparison}(a)). Natural scenes therefore use, on average, a slightly smaller fraction of the available palette and distribute weight more evenly across the colors they do use.

\begin{figure}[t]
    \centering
    \includegraphics[width=0.85\linewidth]{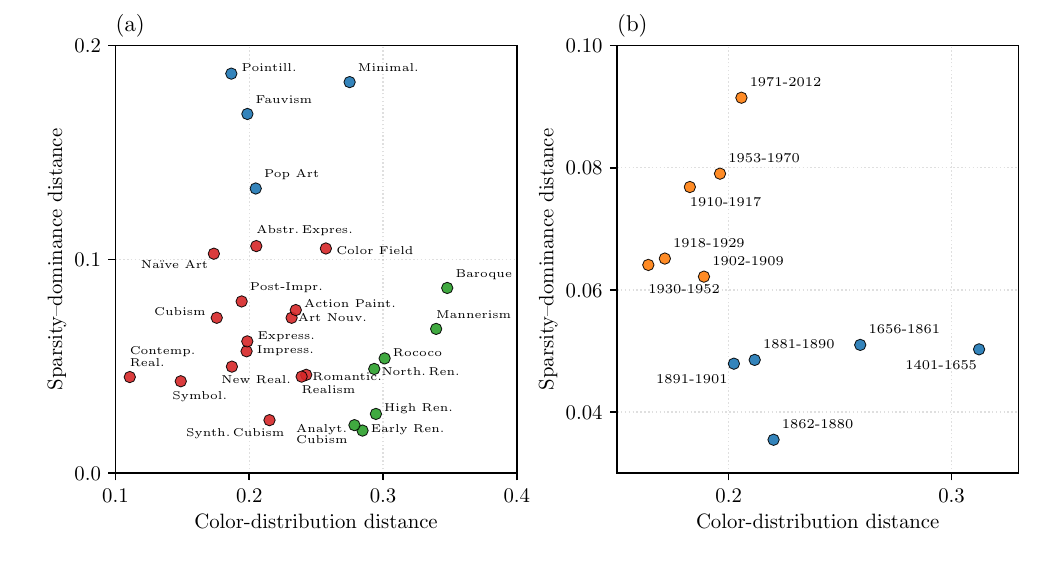}
    \caption{Distance from natural images. Each group is positioned by the Jensen--Shannon distance $D_{\mathrm{col}}$ between its color distribution and that of the natural images (horizontal axis), and by the Euclidean distance $D_{\mathrm{SD}}$ between its mean $(s,d)$ and the natural centroid $(s_{\rm nat},d_{\rm nat})$ (vertical axis). Colors indicate the groups returned by $k$-means on the standardized pair $(D_{\mathrm{col}},D_{\mathrm{SD}})$. (a)~The 26 movements, $k=3$: far in $D_{\mathrm{SD}}$ (blue: Pointillism, Fauvism, Pop Art, Minimalism), far in $D_{\mathrm{col}}$ (green: Renaissance schools, Mannerism, Baroque, Rococo, Analytical Cubism), and the central group (red). (b)~The eleven periods, shown for the $k=2$ partition: 1401--1901 (blue) and 1902--2012 (orange).}
    \label{fig:clustering}
\end{figure}

As in Sec.~\ref{sec:colordistances}, we denote by $D_{\mathrm{col}}$ the Jensen--Shannon distance between the color distribution of a group and that of the natural images. We also define $D_{\mathrm{SD}}$ as the Euclidean distance between their positions in the $(s,d)$ plane. Plotting one against the other separates groups that depart from nature in different ways (Fig.~\ref{fig:clustering}).
We cluster the movements in the plane defined by the two distances $(D_{\mathrm{col}},D_{\mathrm{SD}})$, applying $k$-means to their standardized values. The number of clusters is selected from the elbow of the within-cluster inertia, which yields three groups and remains stable across the search ranges we tested. The silhouette score~\cite{rousseeuw1987silhouettes} provides an independent consistency check, with a local maximum at the same value~\footnote{The search range is set by requiring that no group be reduced to a single member; for the $26$ movements the first singleton appears at $k=8$, so the search runs over $k\le 7$. The elbow returns $k=3$ there and for every range from $k\le 4$ to $k\le 10$. The silhouette at $k=3$ is $0.45$, higher than at any other number of groups up to $k=8$. Its global maximum lies at much larger $k$, since on a sample of this size the silhouette increases almost monotonically toward finer partitions and is maximized by solutions that break the movements into groups of one to three members. We therefore read it as a check on a value selected otherwise, rather than as a criterion for selecting $k$.}.

Interestingly, the resulting three groups (Fig.~\ref{fig:clustering}(a)) do not follow the distinction between figurative and abstract painting, as one may have expected. The largest group, closest to nature along both axes, mixes the two, from Realism, Impressionism, and Post-Impressionism to Cubism, Abstract Expressionism, and Color Field. The second group, comprising Pointillism, Fauvism, Pop Art, and Minimalism, lies far from nature in $D_{\mathrm{SD}}$ but only moderately far in $D_{\mathrm{col}}$. Their positions in Fig.~\ref{fig:SD_comparison}(b) clarify the nature of this departure. Pointillism and Fauvism ($s=0.29$ and $0.31$) use a larger fraction of the palette than natural scenes ($s_{\mathrm{nat}}=0.47$), whereas Minimalism ($s=0.62$) uses a smaller one; Pop Art, at $(s,d)=(0.38,0.69)$, combines broad palette use with strong dominance by a few colors. The third group, comprising the Renaissance schools, Mannerism, Baroque, Rococo, and Analytical Cubism, shows the opposite pattern: it lies far from nature in $D_{\mathrm{col}}$ but remains close in $D_{\mathrm{SD}}$. These movements therefore differ from natural scenes mainly in {which} colors they use, rather than in how many they use or how evenly they distribute them. This is consistent with the restricted pigment repertoire and tonal construction of pre-modern painting and, for Analytical Cubism, with its deliberately near-monochromatic palette.

This partition is robust to the choice of clustering method. Cutting a Ward agglomerative dendrogram~\cite{ward1963,murtagh2014ward} at the largest gap between successive merge heights, a criterion independent of the $k$-means solution, also yields three groups that coincide for 25 of the 26 movements. The sole exception is Pop Art, which occupies an intermediate position in $D_{\mathrm{SD}}$ between the central group and the other members of the second group. This may reflect the visual language of printed advertising on which Pop Art draws: several distinct flat colors reduce sparsity, while a few of them occupy large areas of the image and raise dominance above the natural value.

For the historical periods, the same criterion initially returns two groups, but this result is not robust to the search range~\footnote{For the eleven periods the admissible range is $k\le 4$, where the elbow returns two groups, the same split that the Ward dendrogram gives when cut at its largest gap, but extending the search to $k\le 6$ or beyond returns three instead. The silhouette is of no help either, its values at $k=2$, $3$ and $4$ spanning only $0.011$.}. Figure~\ref{fig:clustering}(b) is therefore better read as a trajectory than as a set of clusters. From the fifteenth to the end of the nineteenth century, $D_{\mathrm{col}}$ decreases from $0.31$ to $0.20$ while $D_{\mathrm{SD}}$ remains close to $0.05$, indicating that the palette approaches that of natural scenes while retaining a similar organization of colors. From the twentieth century onward, $D_{\mathrm{SD}}$ increases, followed by $D_{\mathrm{col}}$ in the last two periods, so that the most recent periods are the furthest from nature in both respects. This trajectory is consistent with the change in the role of color discussed in Sec.~\ref{sec:colordistances}: painting moves closer to natural color statistics while color primarily serves the depiction of the visible world, and departs from them as color increasingly becomes, in the twentieth century, a pictorial subject in its own right.

\section{Survey of perceived color harmony in the sparsity--dominance plane}
\label{sec:survey}

Here we ask whether sparsity and dominance are also related to aesthetic judgment, namely to how harmonious a set of colors is perceived to be. This cannot be tested directly on paintings, whose appreciation also depends on subject, composition, familiarity, and cultural prestige, none of which can be held fixed. We therefore conduct two surveys using abstract images, in which sparsity and dominance can be controlled independently.

The stimuli are generated from the palettes of five well-known paintings spanning a broad range of periods and styles. From each painting, the generated images inherit only the ranking of the 25 colors, from the most to the least represented. For each painting, we sample the $(s,d)$ plane on a regular $5\times5$ grid and generate one image per cell as a partition of the canvas into non-overlapping rectangles. To verify that the results do not depend on this geometry, we run a second survey that is identical in every other respect, but in which the canvas is instead covered by overlapping circular patches (see Figs.~\ref{fig:survey_imagesP1} to \ref{fig:survey_imagesP5} in  Supplemental Information~\ref{subsec:surveydesign}).
Each question displays five images in which one coordinate is fixed while the other spans its full range, and participants are asked to select the one with the \textit{most harmonious color balance}. The construction of the stimuli and the design of the surveys are detailed in Supplemental Information~\ref{subsec:surveydesign}. The surveys were administered through the Qualtrics platform~\cite{qualtrics}. The first collected responses from 385 participants and the second from 606. Participants were recruited among colleagues at \'Ecole Polytechnique and CFM (France) and among volunteers across Europe, without financial incentive; the second survey was also circulated on LinkedIn to broaden the pool further. The sample includes male and female participants aged between 20 and 65, mostly highly educated but from diverse academic backgrounds. Each survey closed with a color-vision screening question; the fraction of participants flagged was negligible, and excluding them leaves all results unchanged.

\begin{figure}[t]
    \centering
    \includegraphics[width=0.85\linewidth]{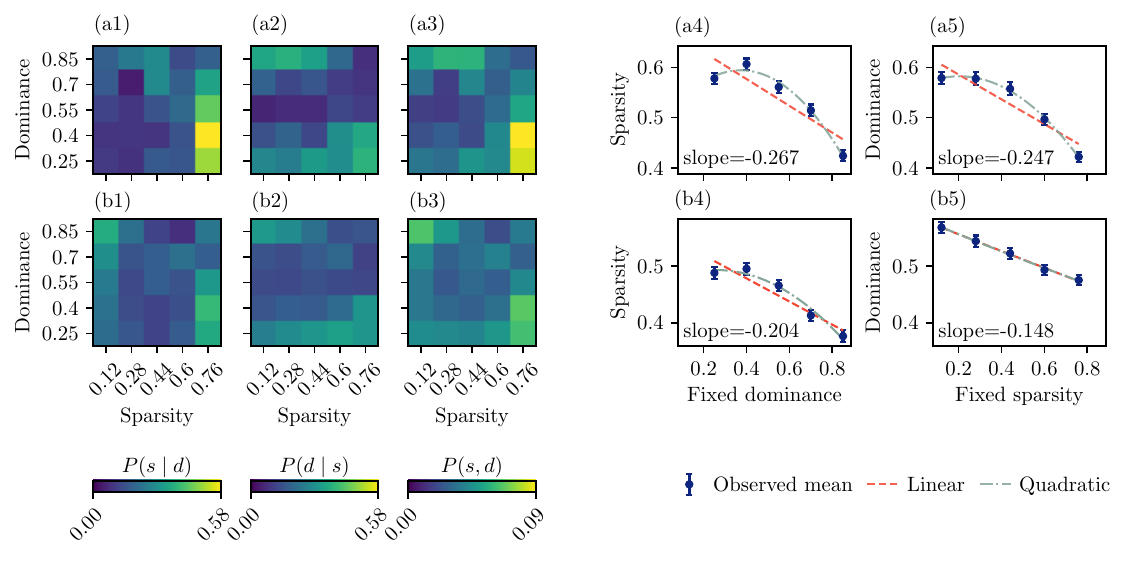}
    \caption{
Survey results. Top row: first survey (non-overlapping rectangles, 409 participants); bottom row: second survey (overlapping circles, 547 participants). (a1,b1)~Distribution of chosen sparsity at fixed dominance, $P(s\,|\,d)$, normalized by rows. (a2,b2)~Distribution of chosen dominance at fixed sparsity, $P(d\,|\,s)$, normalized by columns. (a3,b3)~Overall selection rate $P(s,d)$. (a4,b4)~Mean chosen sparsity at each fixed value of dominance. (a5,b5)~Mean chosen dominance at each fixed value of sparsity. Dashed and dash-dotted lines show linear and quadratic fits, respectively; the slope of the linear fit is reported in each panel.
    }
    \label{fig:survey_summary}
\end{figure}

The results of both surveys are shown in Fig.~\ref{fig:survey_summary}, for rectangular patches (top row) and overlapping circles (bottom row). Since each question displays one row or one column of the $5\times5$ grid, panels (a1), (b1), (a2), and (b2) report the raw responses as conditional probabilities. Panels (a1) and (b1) show the distribution of the sparsity chosen at fixed dominance, $P(s\,|\,d)$, and panels (a2) and (b2) that of the dominance chosen at fixed sparsity, $P(d\,|\,s)$, while panels (a3) and (b3) show the overall selection rate of each pair $(s,d)$. Panels (a4), (b4), (a5), and (b5) report the mean value chosen for one coordinate at each fixed value of the other.

The selection probabilities are far from uniform~\footnote{A $\chi^2$ test against uniformity within each of the ten experimental conditions, five at fixed dominance and five at fixed sparsity, rejects in every one at $p<10^{-6}$, with $\chi^2$ between 34 and 146 for four degrees of freedom and effect sizes $w$ between $0.24$ and $0.50$. Every participant answers each condition exactly once, so the responses entering a given test come from distinct participants and no correction for repeated measures is required.}, and the two surveys, despite using different stimulus geometries and partly different participant pools, display a consistent structure. Choices concentrate around two poles at opposite corners of the plane: low sparsity with high dominance, and high sparsity with low dominance. In other words, participants tend to prefer either many colors arranged in a strong hierarchy or few colors used more evenly. Combinations in which both coordinates are near the middle of their ranges are selected least often. The two remaining corners are also disfavored, corresponding to images that appear almost monochromatic at high sparsity and high dominance, or comparatively unordered at low sparsity and low dominance. Harmony therefore appears to depend on the combination of the two coordinates rather than on either one alone.
The two surveys differ mainly in the relative weight of the two preferred poles. With rectangular patches, the high-sparsity, low-dominance pole is selected more often, so that a small palette used evenly is preferred to a larger one with a strong hierarchy. With overlapping circles, the two poles are selected at similar rates. The change in geometry therefore affects the balance between the two poles, but not the two-pole structure itself.
In all four cases shown in the right part of  Fig.~\ref{fig:survey_summary} (panels (a4), (a5), (b4), and (b5)), the mean chosen value of one coordinate decreases as the fixed value of the other increases. A higher value of one coordinate is therefore compensated by a lower value of the other. Because choices within each condition tend to accumulate toward the ends of the available range, however, the mean often falls in a region that is itself rarely selected. These slopes therefore reveal the direction of the compensation, but should not be interpreted as measuring its strength.
Taken together, these results suggest that judgments of harmonious color balance share a common component across participants, in which the number of colors and the evenness of their use compensate each other.

Perhaps surprisingly, this bimodal structure of preference has no direct counterpart in the statistics of either paintings or natural images. This contrasts with our previous results, where preferred hue separations closely matched those most frequently found in natural landscapes~\cite{forni2026harmonious}, and preferred structural complexity similarly aligned with that of natural images~\cite{Lakhal_2020}. Here, by contrast, aesthetic preference in the sparsity--dominance plane does not simply reproduce the distribution of naturally occurring or historically realized color organizations. This suggests that the appreciation of color balance depends on additional aspects of the image beyond the global statistics captured by sparsity and dominance. In paintings, such effects may include subject matter, composition, familiarity, and cultural context, which could substantially reshape preferences once color is embedded in a complete pictorial scene.

\section{Concluding remarks} 
\label{sec:conclusion}

Let us summarize the main contributions of this work. We represented 69{,}486 Western paintings by their distributions over a fixed palette of 25 colors in the perceptually uniform OKLab space. Studying the temporal evolution of individual colors, we found that the chromatic history of Western painting unfolds through several shifts staggered in time: dark earths recede throughout, muted mid-tones rise before about 1880 and then level off, and saturated hues increase mainly in the twentieth century. At the level of complete color distributions, historical periods drift gradually and cumulatively, whereas artistic movements form blocks that chronology organizes only partially. Up to Realism, movements remain chromatically very similar; in the twentieth century, by contrast, they become strongly differentiated, as color increasingly becomes part of each movement's signature.
We then introduced two interpretable descriptors, sparsity and dominance, which reduce the color distribution of a painting to how many colors are used and how unevenly they are represented. In this plane, the historical periods trace a two-phase trajectory: a widening of the palette at roughly constant dominance until the 1910s, followed by a marked rise in dominance after 1950. Combined with the Jensen--Shannon distance, these descriptors distinguish two ways in which paintings depart from natural scenes: through the colors themselves, or through how broadly and evenly they are used. This distinction does not follow the conventional divide between figurative and abstract painting. Over time, painting first moves closer to natural color statistics and then away from them during the twentieth century, consistently with the changing role of color documented throughout the paper.
Finally, two surveys show that perceived color harmony can also be organized in terms of sparsity and dominance. Preferences concentrate around two complementary regimes: many colors arranged in a strong hierarchy, or few colors used more evenly. The two coordinates therefore compensate one another, suggesting that judgments of harmonious color balance are not controlled by palette size or dominance alone, but by their combination. Unlike our previous findings for preferred hue separations and structural complexity~\cite{forni2026harmonious,Lakhal_2020}, however, this preference structure is not mirrored in the statistics of either paintings or natural images, suggesting that color harmony in complete artworks may depend importantly on pictorial and contextual factors beyond these global color statistics.

Our analysis naturally comes with a few qualifications. First, it relies on digital reproductions, so the measured colors inevitably reflect not only the paintings themselves but also the conditions under which they were digitized, including lighting, white balance, compression, and, for older works, the aging of varnishes and pigments. Such effects may contribute to some of the differences between early and recent works. 
The effect of an aged varnish, in particular, has been shown to differ significantly from one painting to another~\cite{trumpy2015experimental, linhares2020chromatic, kirchner2018digitally}, and can only be corrected through hyperspectral measurements. No clear general aging trends seem to exist, making it difficult to account for in our large scale analysis.
 These factors are, however, unlikely to account on their own for the structure observed here: the evolution is not a simple monotonic function of age, but involves distinct groups of colors changing at different times, a two-stage trajectory in the sparsity--dominance plane, and substantial differences between movements of similar periods. Second, the dated corpus is unevenly distributed in time, with most works concentrated in the nineteenth and twentieth centuries. As a result, the earliest periods necessarily cover much broader time intervals, and the pre-modern evolution is resolved more coarsely than the modern one. Finally, our representation deliberately abstracts away from spatial organization: a color histogram captures which colors are used and in what proportions, but not where they appear or how they interact locally, even though chromatic context is known to affect perception~\cite{shevell2008color}.

These considerations point to several directions for future work. A first is to complement color distributions with spatial descriptors, for instance by combining sparsity and dominance with measures of structural complexity~\cite{sigaki2018history}. Such descriptors would make it possible to distinguish, for example, whether a dominant color is spread across the canvas or concentrated in a single region, and more generally how colors present in different proportions are arranged in space.
A second direction is to quantify more directly the contribution of material aging and digitization to the measured color statistics. Comparisons of works before and after restoration, or with reconstructions of their original colors based on pigment analysis, could provide an independent estimate of the magnitude of these effects. Hyperspectral measurements, already applied to paintings~\cite{montagner2016statistics}, could similarly be used on a subset of the corpus to compare digital reproductions with the colors of the physical works. Such analyses would help separate these measurement and conservation effects from the historical changes in color usage identified here.
A third direction is to extend the analysis beyond Western painting, beginning with the Ukiyo-e prints excluded here. Artistic traditions with different histories of pigments and materials would provide a natural test of which trends are specific to Western painting and which reflect more general constraints on color use.
Finally, larger and more diverse survey panels would make it possible to test whether the two-pole structure of preferences persists across cultures, ages, and levels of artistic training. Since this structure is already robust to a change in stimulus geometry, a further step would be to use stimuli that preserve the spatial organization of real paintings and test whether the same pattern survives when color is embedded in a full composition.

\section{Acknowledgments}
We are deeply grateful to all the participants of our surveys, whose contribution was essential to this work. We also thank Damien Challet and Costanza Maria Morcella for fruitful discussions. This research was conducted within the Econophysics \& Complex Systems Research Chair, under the aegis of the Fondation du Risque and Capital Fund Management.

\bibliography{refs}

@misc{qualtrics,
  title  = {Qualtrics XM},
  howpublished = {\url{https://www.qualtrics.com/}}
}

@article{lin1991,
  author  = {Lin, Jianhua},
  title   = {Divergence Measures Based on the Shannon Entropy},
  journal = {IEEE Transactions on Information Theory},
  volume  = {37},
  number  = {1},
  pages   = {145--151},
  year    = {1991},
  doi     = {10.1109/18.61115}
}

@misc{kaggle_12000,
  title  = {Kaggle Landscape Recognition Image Database (12000 Images)},
  year   = {2024},
  howpublished = {\url{https://www.kaggle.com/datasets/utkarshsaxenadn/landscape-recognition-image-dataset-12k-images}}
}

@article{Lakhal_2020,
   title={Beauty and structural complexity},
   volume={2},
   ISSN={2643-1564},
   number={2},
   journal={Physical Review Research},
   publisher={American Physical Society (APS)},
   author={Lakhal, Samy and Darmon, Alexandre and Bouchaud, Jean-Philippe and Benzaquen, Michael},
   year={2020}, 
   month=jun }

@misc{ottosson2020perceptual,
  author       = {Ottosson, Bj{\"o}rn},
  title        = {A perceptual color space for image processing},
  year         = {2020},
  howpublished = {\url{https://bottosson.github.io/posts/oklab/}}
}

@inproceedings{anderson1996proposal,
  title={Proposal for a standard default color space for the internet—srgb},
  author={Anderson, Matthew and Motta, Ricardo and Chandrasekar, Srinivasan and Stokes, Michael},
  booktitle={Color and imaging conference},
  volume={4},
  pages={238--245},
  year={1996},
  organization={Society of Imaging Science and Technology}
}

@inproceedings{susstrunk1999standard,
  title={Standard RGB color spaces},
  author={S{\"u}sstrunk, Sabine and Buckley, Robert and Swen, Steve},
  booktitle={Color and imaging conference},
  volume={7},
  pages={127--134},
  year={1999},
  organization={Society of Imaging Science and Technology}
}

@article{Histoire_des_couleur_Pastoureau,
 ISSN = {00462616, 21010064},
 author = {Michel Pastoureau},
 journal = {Ethnologie française},
 number = {4},
 pages = {368--377},
 publisher = {Presses Universitaires de France},
 title = {Une histoire des couleurs est-elle possible?},
 urldate = {2026-01-05},
 volume = {20},
 year = {1990}
}

@book{pastoureau2005petit,
    author = {Pastoureau, Michel and Simonnet, Dominique},
    title = {Le petit livre des couleurs},
    publisher = {(No Title)},
    year = {2005}
}

@article{redies2025toolbox,
  title={A toolbox for calculating quantitative image properties in aesthetics research},
  author={Redies, Christoph and Bartho, Ralf and Ko{\ss}mann, Lisa and Spehar, Branka and H{\"u}bner, Ronald and Wagemans, Johan and Hayn-Leichsenring, Gregor U},
  journal={Behavior Research Methods},
  volume={57},
  number={4},
  pages={117},
  year={2025},
  publisher={Springer}
}

@article{sigaki2018history,
  title={History of art paintings through the lens of entropy and complexity},
  author={Sigaki, Higor YD and Perc, Matja{\v{z}} and Ribeiro, Haroldo V},
  journal={Proceedings of the National Academy of Sciences},
  volume={115},
  number={37},
  pages={E8585--E8594},
  year={2018},
  publisher={National Academy of Sciences}
}

@article{kim2014large,
  title={Large-scale quantitative analysis of painting arts},
  author={Kim, Daniel and Son, Seung-Woo and Jeong, Hawoong},
  journal={Scientific reports},
  volume={4},
  number={1},
  pages={7370},
  year={2014},
  publisher={Nature Publishing Group UK London}
}

@article{montagner2016statistics,
  title={Statistics of colors in paintings and natural scenes},
  author={Montagner, Cristina and Linhares, {Jo{\~a}o} M. M. and Vilarigues, M{\'a}rcia and Nascimento, S{\'e}rgio MC},
  journal={Journal of the Optical Society of America A},
  volume={33},
  number={3},
  pages={A170--A177},
  year={2016},
  publisher={OSA}
}

@book{wolfflin2012principles,
  title={Principles of art history},
  author={W{\"o}lfflin, Heinrich},
  year={2012},
  publisher={Courier Corporation}
}

@article{nascimento2021naturalness,
  title={Naturalness and aesthetics of colors--Preference for color compositions perceived as natural},
  author={Nascimento, S{\'e}rgio MC and Albers, Anke Marit and Gegenfurtner, Karl R},
  journal={Vision Research},
  volume={185},
  pages={98--110},
  year={2021},
  publisher={Elsevier}
}

@article{nakauchi2022regularity,
  title={Regularity of colour statistics in explaining colour composition preferences in art paintings},
  author={Nakauchi, Shigeki and Tamura, Hideki},
  journal={Scientific Reports},
  volume={12},
  number={1},
  pages={14585},
  year={2022},
  publisher={Nature Publishing Group UK London}
}

@article{Graham2007StatisticalRO,
  title={Statistical regularities of art images and natural scenes: spectra, sparseness and nonlinearities.},
  author={Daniel J. Graham and David J. Field},
  journal={Spatial vision},
  year={2007},
  volume={21 1-2},
  pages={
          149-64
        },
  url={https://api.semanticscholar.org/CorpusID:678769}
}

@misc{wikiart_kaggle,
  title  = {WikiArt Dataset},
  year   = {2022},
  howpublished = {\url{https://www.kaggle.com/datasets/steubk/wikiart}}
}

@book{gage1993colour,
  author    = {Gage, John},
  title     = {Colour and Culture: Practice and Meaning from Antiquity to Abstraction},
  publisher = {Thames and Hudson},
  address   = {London},
  year      = {1993}
}

@article{shevell2008color,
  author  = {Shevell, Steven K. and Kingdom, Frederick A. A.},
  title   = {Color in Complex Scenes},
  journal = {Annual Review of Psychology},
  volume  = {59},
  pages   = {143--166},
  year    = {2008},
  doi     = {10.1146/annurev.psych.59.103006.093619}
}

@article{lee2018heterogeneity,
  author  = {Lee, Byunghwee and Kim, Daniel and Sun, Seunghye and Jeong, Hawoong and Park, Juyong},
  title   = {Heterogeneity in chromatic distance in images and characterization of massive painting data set},
  journal = {PLOS ONE},
  volume  = {13},
  number  = {9},
  pages   = {e0204430},
  year    = {2018},
  doi     = {10.1371/journal.pone.0204430}
}

@article{feitosa2020graffiti,
  author  = {Feitosa-Santana, Claudia and Gaddi, Carlo M. and Gomes, Andreia E. and Nascimento, S\'{e}rgio M. C.},
  title   = {Art through the Colors of Graffiti: From the Perspective of the Chromatic Structure},
  journal = {Sensors},
  volume  = {20},
  number  = {9},
  pages   = {2531},
  year    = {2020},
  doi     = {10.3390/s20092531}
}

@article{taylor1999fractal,
  author  = {Taylor, Richard P. and Micolich, Adam P. and Jonas, David},
  title   = {Fractal analysis of Pollock's drip paintings},
  journal = {Nature},
  volume  = {399},
  pages   = {422},
  year    = {1999}
}

@article{berezhnoy2007vangogh,
  author  = {Berezhnoy, Igor and Postma, Eric and van den Herik, Jaap},
  title   = {Computer analysis of Van Gogh's complementary colours},
  journal = {Pattern Recognition Letters},
  volume  = {28},
  pages   = {703--709},
  year    = {2007},
  doi     = {10.1016/j.patrec.2006.08.002}
}

@article{papia2025rothko,
  author  = {Papia, E.-M. and Kondi, A.},
  title   = {Quantifying subtle color transitions in Mark Rothko's abstract paintings through K-means clustering and Delta E analysis},
  journal = {Journal of Cultural Heritage},
  volume  = {72},
  pages   = {194--204},
  year    = {2025}
}

@inproceedings{icoglu2004classification,
  title={Classification and indexing of paintings based on art movements},
  author={Icoglu, Oguz and Gunsel, Bilge and Sariel, Sanem},
  booktitle={2004 12th European Signal Processing Conference},
  pages={749--752},
  year={2004},
  organization={IEEE}
}

@article{falomir2018qartlearn,
  author  = {Falomir, Zoe and Museros, Lled\'{o} and Sanz, Ismael and Gonzalez-Abril, Luis},
  title   = {Categorizing paintings in art styles based on qualitative color descriptors, quantitative global features and machine learning (QArt-Learn)},
  journal = {Expert Systems with Applications},
  volume  = {97},
  pages   = {83--94},
  year    = {2018},
  doi     = {10.1016/j.eswa.2017.11.056}
}

@inproceedings{elgammal2018shape,
  author    = {Elgammal, Ahmed and Liu, Bingchen and Kim, Diana and Elhoseiny, Mohamed and Mazzone, Marian},
  title     = {The Shape of Art History in the Eyes of the Machine},
  booktitle = {Proceedings of the Thirty-Second AAAI Conference on Artificial Intelligence (AAAI-18)},
  year      = {2018}
}

@misc{honna2024ukiyoe,
  author = {Honna, Shinichi and Matsui, Akira},
  title  = {A Deep Learning Approach to Quantitative Analysis of Japanese Artworks: Exploring the Latent Features of Ukiyo-e Artists},
  year   = {2024},
  note   = {SSRN preprint 4706101}
}

@article{sartori2015affective,
  author  = {Sartori, Andreza and Yanulevskaya, Victoria and Akdag Salah, Almila and Uijlings, Jasper and Bruni, Elia and Sebe, Nicu},
  title   = {Affective analysis of professional and amateur abstract paintings using statistical analysis and art theory},
  journal = {ACM Transactions on Interactive Intelligent Systems},
  volume  = {5},
  number  = {2},
  pages   = {8},
  year    = {2015},
  doi     = {10.1145/2768209}
}

@article{chen2024emotion,
  author  = {Chen, Chin-Ling and Huang, Qing-Yang and Zhou, Ming and Huang, Der-Chen and Liu, Ling-Chun and Deng, Yong-Yuan},
  title   = {Quantified emotion analysis based on design principles of color feature recognition in pictures},
  journal = {Multimedia Tools and Applications},
  volume  = {83},
  pages   = {57243--57267},
  year    = {2024},
  doi     = {10.1007/s11042-023-17286-6}
}

@article{borisov2024pricing,
  title={How does the color palette affect the pricing of abstract paintings?},
  author={Borisov, Maksim and Kolycheva, Valeria and Semenov, Alexander and Grigoriev, Dmitry},
  journal={Information Systems Frontiers},
  pages={1--23},
  year={2025},
  publisher={Springer}
}

@article{stepanova2019palettes,
  author  = {Stepanova, Elena},
  title   = {The impact of color palettes on the prices of paintings},
  journal = {Empirical Economics},
  volume  = {56},
  number  = {2},
  pages   = {755--773},
  year    = {2019},
  doi     = {10.1007/s00181-017-1413-4}
}

@techreport{gorin2025stateoftheart,
  author      = {Gorin, Cl\'{e}ment and Heblich, Stephan and Zylberberg, Yanos},
  title       = {State of the Art: Economic Development Through the Lens of Paintings},
  institution = {National Bureau of Economic Research},
  type        = {NBER Working Paper},
  number      = {33976},
  year        = {2025},
  note        = {Revised June 2026}
}

@article{albright2023pollution,
  author  = {Albright, Anna Lea and Huybers, Peter},
  title   = {Paintings by Turner and Monet depict trends in 19th century air pollution},
  journal = {Proceedings of the National Academy of Sciences},
  volume  = {120},
  number  = {6},
  pages   = {e2219118120},
  year    = {2023},
  doi     = {10.1073/pnas.2219118120}
}

@article{nascimento2017colors,
  author  = {Nascimento, S\'{e}rgio M. C. and Linhares, {Jo\~{a}o} M. M. and Montagner, Cristina and Jo\~{a}o, Catarina A. R. and Amano, Kinjiro and Alfaro, Catarina and Bail\~{a}o, Ana},
  title   = {The colors of paintings and viewers' preferences},
  journal = {Vision Research},
  volume  = {130},
  pages   = {76--84},
  year    = {2017},
  doi     = {10.1016/j.visres.2016.11.006}
}

@article{albers2020independent,
  author  = {Albers, Anke Marit and Gegenfurtner, Karl R. and Nascimento, S\'{e}rgio M. C.},
  title   = {An independent contribution of colour to the aesthetic preference for paintings},
  journal = {Vision Research},
  volume  = {177},
  pages   = {109--117},
  year    = {2020},
  doi     = {10.1016/j.visres.2020.08.005}
}

@article{forni2026harmonious,
  author  = {Forni, Ortensia and Darmon, Alexandre and Benzaquen, Michael},
  title   = {Harmonious color pairings: Insights from human preference and natural hue statistics},
  journal = {iScience},
  volume  = {29},
  pages   = {116038},
  year    = {2026},
  doi     = {10.1016/j.isci.2026.116038}
}

@article{ward1963,
  author  = {Joe H. Ward},
  title   = {Hierarchical Grouping to Optimize an Objective Function},
  journal = {Journal of the American Statistical Association},
  volume  = {58},
  number  = {301},
  pages   = {236--244},
  year    = {1963},
  doi     = {10.1080/01621459.1963.10500845},
}

@article{murtagh2014ward,
  author  = {Fionn Murtagh and Pierre Legendre},
  title   = {Ward's Hierarchical Agglomerative Clustering Method:
             Which Algorithms Implement Ward's Criterion?},
  journal = {Journal of Classification},
  volume  = {31},
  number  = {3},
  pages   = {274--295},
  year    = {2014},
  doi     = {10.1007/s00357-014-9161-z},
}

@article{rousseeuw1987silhouettes,
  title={Silhouettes: a graphical aid to the interpretation and validation of cluster analysis},
  author={Rousseeuw, Peter J},
  journal={Journal of computational and applied mathematics},
  volume={20},
  pages={53--65},
  year={1987},
  publisher={Elsevier}
}

@article{damgaard2000describing,
  title={Describing inequality in plant size or fecundity},
  author={Damgaard, Christian and Weiner, Jacob},
  journal={Ecology},
  volume={81},
  number={4},
  pages={1139--1142},
  year={2000},
  publisher={Wiley Online Library}
}

@book{golding1961cubism,
  author    = {Golding, John},
  title     = {Cubism: A History and an Analysis, 1907--1914},
  publisher = {George Wittenborn},
  address   = {New York},
  year      = {1959}
}

@book{elderfield1976wild,
  title={The''Wild Beasts''; Fauvism and its affinities},
  author={Elderfield, John},
  year={1976}
}

@book{philipp2003bright,
  title={Bright Earth: Art and The Invention of Color},
  author={Ball, Philip},
  year={2003},
  publisher={Chicago, University of Chicago Press}
}

@article{linhares2020chromatic,
  title={Chromatic changes in paintings of Adriano de Sousa Lopes after the removal of aged varnish},
  author={Linhares, Jo{\~a}o and Cardeira, Liliana and Bail{\~a}o, Ana and Pastilha, Ruben and Nascimento, S{\'e}rgio},
  journal={Conservar Patrim{\'o}nio},
  volume={34},
  pages={50--64},
  year={2020}
}

@article{trumpy2015experimental,
  title={Experimental study on merits of virtual cleaning of paintings with aged varnish},
  author={Trumpy, Giorgio and Conover, Damon and Simonot, Lionel and Thoury, Mathieu and Picollo, Marcello and Delaney, John K},
  journal={Optics express},
  volume={23},
  number={26},
  pages={33836--33848},
  year={2015},
  publisher={Optical Society of America}
}

@article{kirchner2018digitally,
  title={Digitally reconstructing van Gogh's field with irises near Arles. Part 1: varnish},
  author={Kirchner, Eric and van der Lans, Ivo and Ligterink, Frank and Hendriks, Ella and Delaney, John},
  journal={Color Research \& Application},
  volume={43},
  number={2},
  pages={150--157},
  year={2018},
  publisher={Wiley Online Library}
}

\clearpage
\section{Supplemental information}

\subsection{Additional figures}
\begin{figure}[H]
    \centering
    \includegraphics[width=0.85\linewidth]{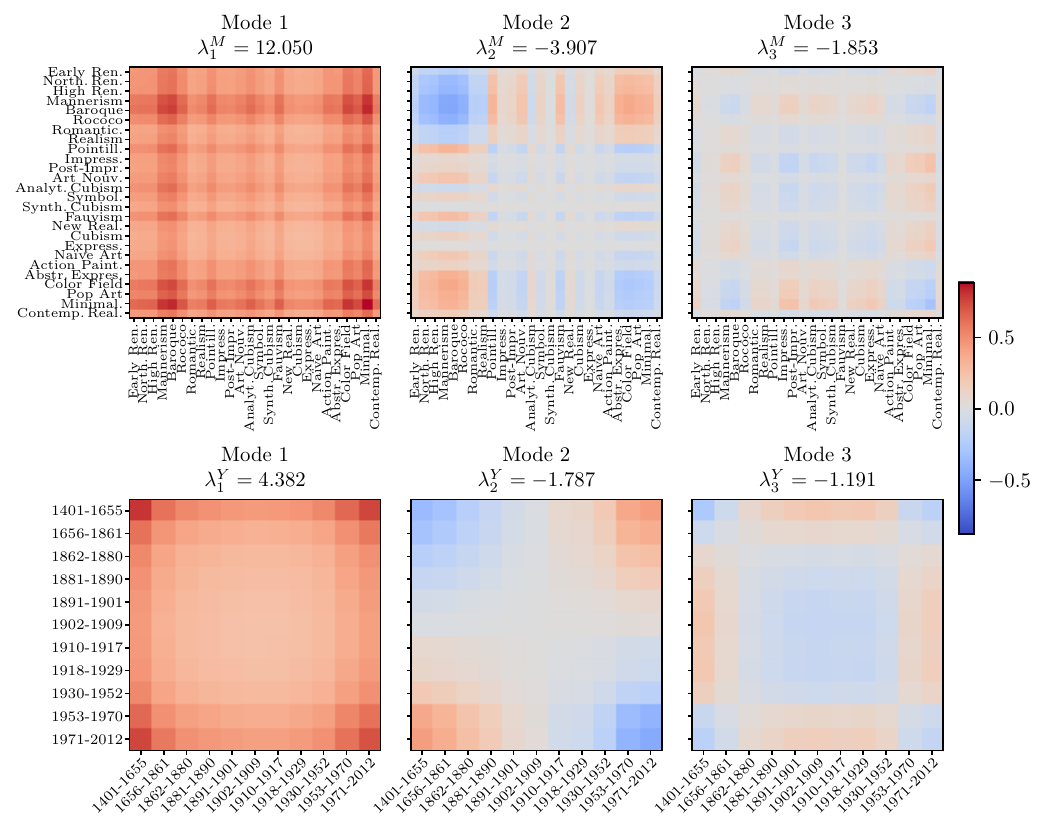}
    \caption{Spectral decomposition of the Jensen--Shannon distance matrices of Fig.~\ref{fig:matrices_JS_colors_movs_years}, for the 26 movements (top row) and the eleven periods (bottom row). Each panel shows the contribution $\lambda_i \mathbf{v}_i \mathbf{v}_i^{\top}$ of one of the three modes with the largest absolute eigenvalues $\lambda_i$, $\mathbf{v}_i$ being the corresponding eigenvector. The eigenvalues are reported above each panel, and the color scale is common to all panels.}
    \label{fig:matrix_colors_pca}
\end{figure}

\clearpage

\begin{figure}[H]
    \centering
    \includegraphics[width=0.85\linewidth]{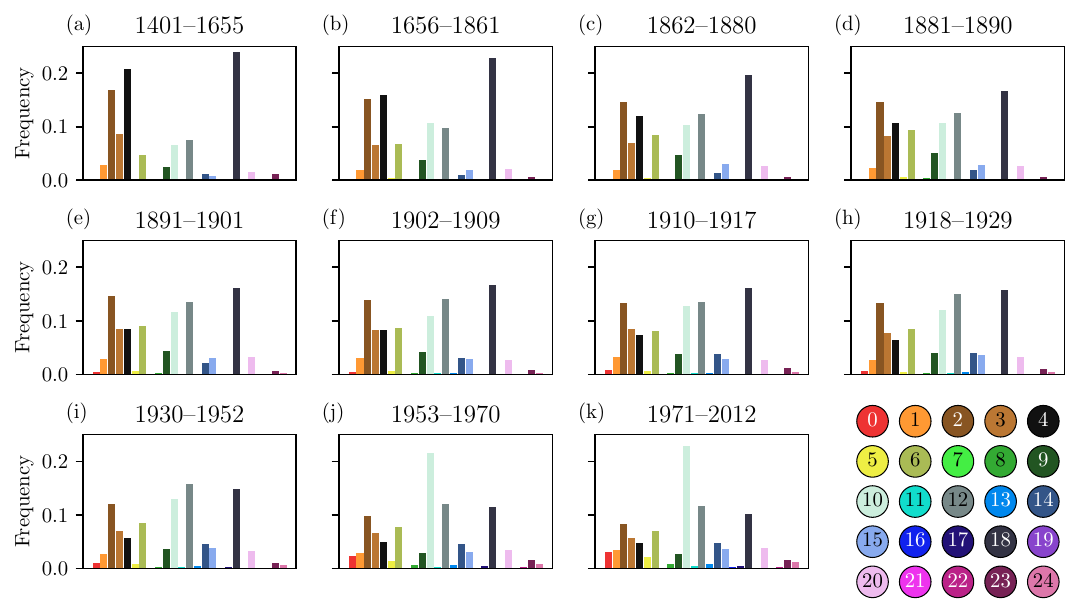}
    \caption{Color distributions $p_t(k)$ of the eleven periods, corresponding to the columns of Fig.~\ref{fig:heatmap_zscore_clusters}(a). Each bar gives the frequency of one of the 25 reference colors and is drawn in the color it represents. The last panel shows the 25 colors of the palette, indexed from 0 to 24.}
\label{fig:color_distribution_time_windows}
\end{figure}
 
\subsection{Reference palette}
\label{sec:paletteconstruction}
Since our data are digital images, displayed on screens in the standard sRGB space~\cite{anderson1996proposal, susstrunk1999standard}, we draw the candidate colors from a regular $16\times16\times16$ grid of the sRGB cube---$4096$ evenly spaced candidates---and convert it to OKLab. We then select $25$ of these candidates by a \emph{maximin} rule: the first color is the candidate nearest neutral mid-gray, and each subsequent color is the one that maximizes its minimum OKLab distance to those already chosen, so that the selected set is spread as widely as possible over the perceptual space. Because the maximin rule gravitates toward the extreme vertices of the sRGB gamut, we further require each selected color to lie at least 0.015 (in OKLab units) inside the gamut boundary: every reference color acts as the prototype of the pixels assigned to it, and an interior anchor represents them with less bias than one located on the boundary. The procedure yields a perceptually well-spread palette spanning warm and cool hues, neutrals, and a black-to-white value axis without privileging any region.

\subsection{Lorenz asymmetry coefficient}
\label{sec:lorenz}

\begin{figure}[H]
    \centering
    \includegraphics[width=0.42\linewidth]{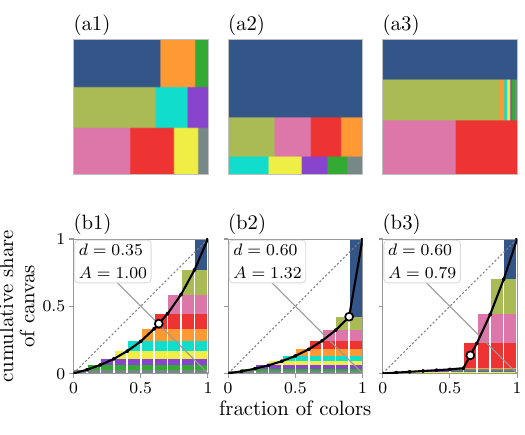}
    \caption{Same increase in dominance with two different Lorenz curves. (a1)--(a3)~Schematic canvases built from ten colors of the palette. (b1)--(b3)~Corresponding Lorenz curves, with the colors sorted by increasing proportion, and values of the dominance $d$ and of the Lorenz asymmetry coefficient $A$. The dotted line is the line of equality and the solid gray line is the anti-diagonal $x+y=1$. $A$ is the sum of the coordinates of the white dot (Sec.~\ref{sec:lorenz}) and is larger (smaller) than 1 when the dot lies above (below) the gray line. Starting from (a1), dominance increases by the same amount when one color takes over~(a2) and when six colors become marginal~(a3), while $A$ increases in the first case and decreases in the second.}
    \label{fig:schematic_DvsA}
\end{figure}

The Lorenz asymmetry coefficient $A$~\cite{damgaard2000describing} distinguishes palettes of equal dominance on the basis of how the canvas is shared among their colors. The same dominance can arise because a few colors cover much more than an equal share of the canvas, or because a few colors cover much less. $A$ is larger than 1 in the first case, smaller than 1 in the second, and equal to 1 when the Lorenz curve is symmetric (Fig.~\ref{fig:schematic_DvsA}).\\
The Lorenz curve plots the cumulative share of the canvas $\sum_{j\le i} p_j$ covered by the $i$ least used colors against the fraction $i/N_{\mathrm{act}}$ of active colors they represent, with $i = 0,\dots,N_{\mathrm{act}}$ and $p_1 \le \dots \le p_{N_{\mathrm{act}}}$ the proportions introduced in Sec.~\ref{sec:sd}. The curve is flatter than the line of equality over the colors covering less than the equal share $\mu = 1/N_{\mathrm{act}}$ and steeper over the others. $A$ is the sum of the two coordinates of the point where the curve passes from one regime to the other, shown as a white dot in Fig.~\ref{fig:schematic_DvsA}.
Denoting by $m$ the number of colors with $p_i < \mu$, the transition occurs between the $m$-th and $(m+1)$-th colors, and we locate it by linear interpolation at the abscissa $(m+\delta)/N_{\mathrm{act}}$, with
\begin{equation}
\delta = \frac{\mu - p_m}{p_{m+1} - p_m},
\label{eq:delta}
\end{equation}
so that $A$ reads
\begin{equation}
A = \frac{m+\delta}{N_{\mathrm{act}}} + \sum_{i=1}^{m} p_i + \delta\, p_{m+1}.
\label{eq:lorenz_asym}
\end{equation}


\begin{figure}[H]
    \centering
    \includegraphics[width=0.85\linewidth]{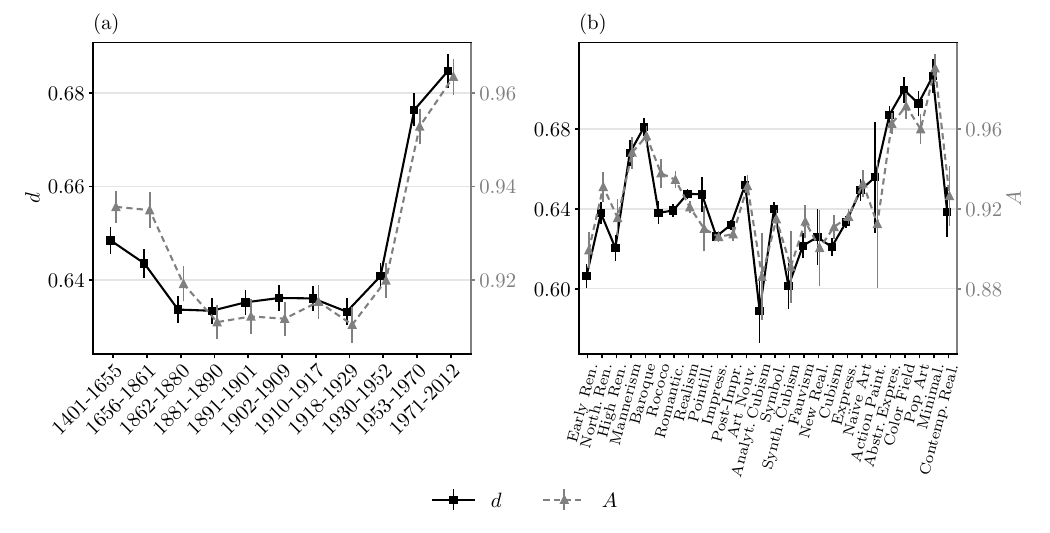}
    \caption{Mean dominance $d$ and mean Lorenz asymmetry coefficient $A$ for (a)~historical periods and (b)~artistic movements ordered by median year. $d$ is read on the left axis and $A$ on the right axis, both with the same unit length. Error bars are 95\% confidence intervals of the mean.}
    \label{fig:DA}
\end{figure}

\subsection{Survey design}
\label{subsec:surveydesign}
\paragraph{Stimuli.} The stimuli were built from the palettes of five well-known paintings, chosen to span a broad range of periods and styles. We chose Raphael's \emph{The Triumph of Galatea} (1512), Turner's \emph{The Burning of the Houses of Parliament} (1834), C\'ezanne's \emph{Mont Sainte-Victoire} (1885), Van Gogh's \emph{The Starry Night} (1889) and Rothko's \emph{Multiform} (1948). For each painting we compute the 25-value color distribution as in Sec.~\ref{sec:data} and rank the 25 colors by decreasing weight. This ranking is the only property the generated images inherit from their source. Every image was generated on a $100\times100$ pixel canvas, matching the $10^4$ pixels sampled from each artwork (Sec.~\ref{sec:data}), so that sparsity and dominance remain directly comparable between generated images and paintings. We sample the $(s,d)$ plane on a regular $5\times5$ grid, with $s\in\{0.12,0.28,0.44,0.60,0.76\}$ and $d\in\{0.25,0.40,0.55,0.70,0.85\}$, giving 25 target cells per painting. Since $s=1-N_{\mathrm{act}}/25$, the sparsity values correspond to images built from exactly $N_{\mathrm{act}}=22$, 18, 14, 10 and 6 active colors, so that each step along the horizontal axis removes four colors from the palette. Within each cell, the proportions of the $N_{\mathrm{act}}$ active colors are chosen so that their Gini coefficient reproduces the target dominance. The value $d=0.85$ keeps the grid evenly spaced but cannot be reached for $N_{\mathrm{act}}=6$, since the Gini coefficient of six colors cannot exceed $1-1/6\simeq0.83$ [Eq.~(\ref{eq:Dmax})]. In the cell $(s,d)=(0.76,0.85)$, the images are therefore generated with the dominance closest to the target, $d\simeq0.83$.\\

\begin{figure}[H]
    \centering
    \includegraphics[width=\linewidth]{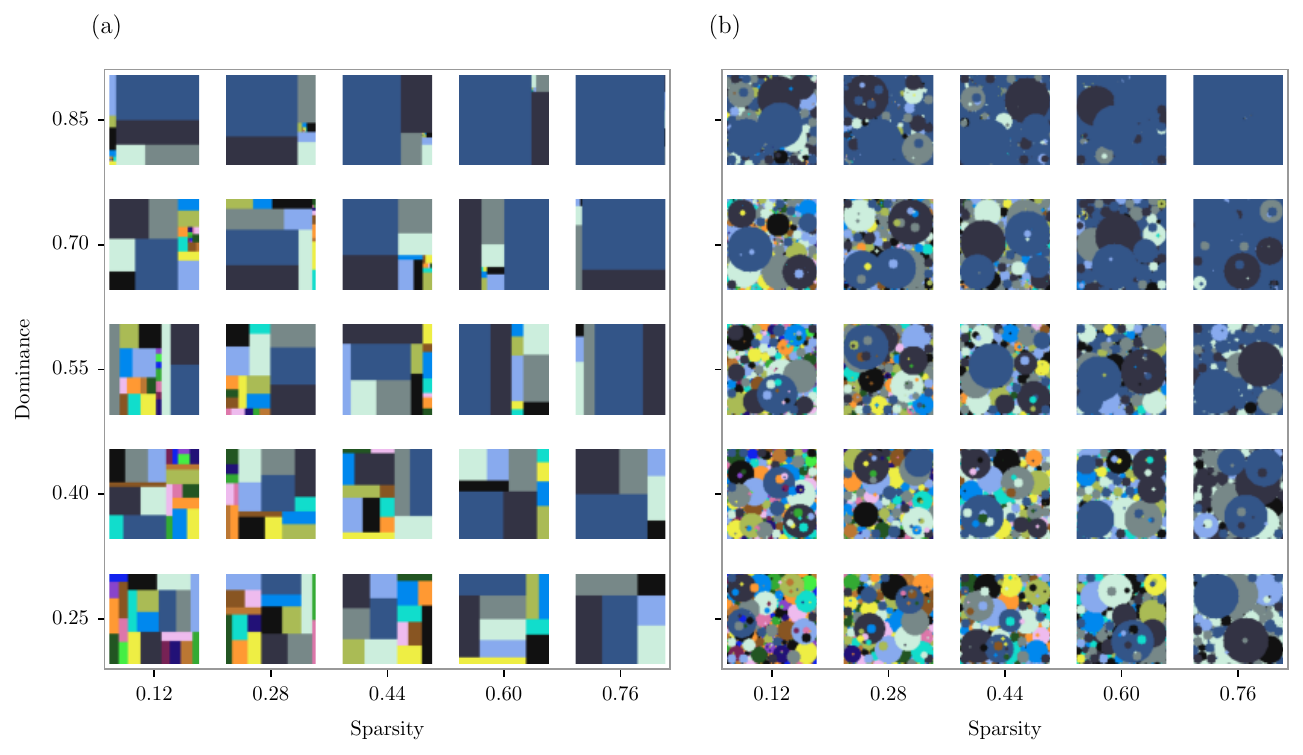}
    \caption{Stimuli generated from the palette of Van Gogh's \emph{The Starry Night} (1889), for the first survey with non-overlapping rectangular patches (a) and for the second survey with overlapping circular patches (b). Each image corresponds to one cell of the $5\times5$ grid in the $(s,d)$ plane.}
    \label{fig:survey_imagesP1}
\end{figure}

\begin{figure}[H]
    \centering
    \includegraphics[width=\linewidth]{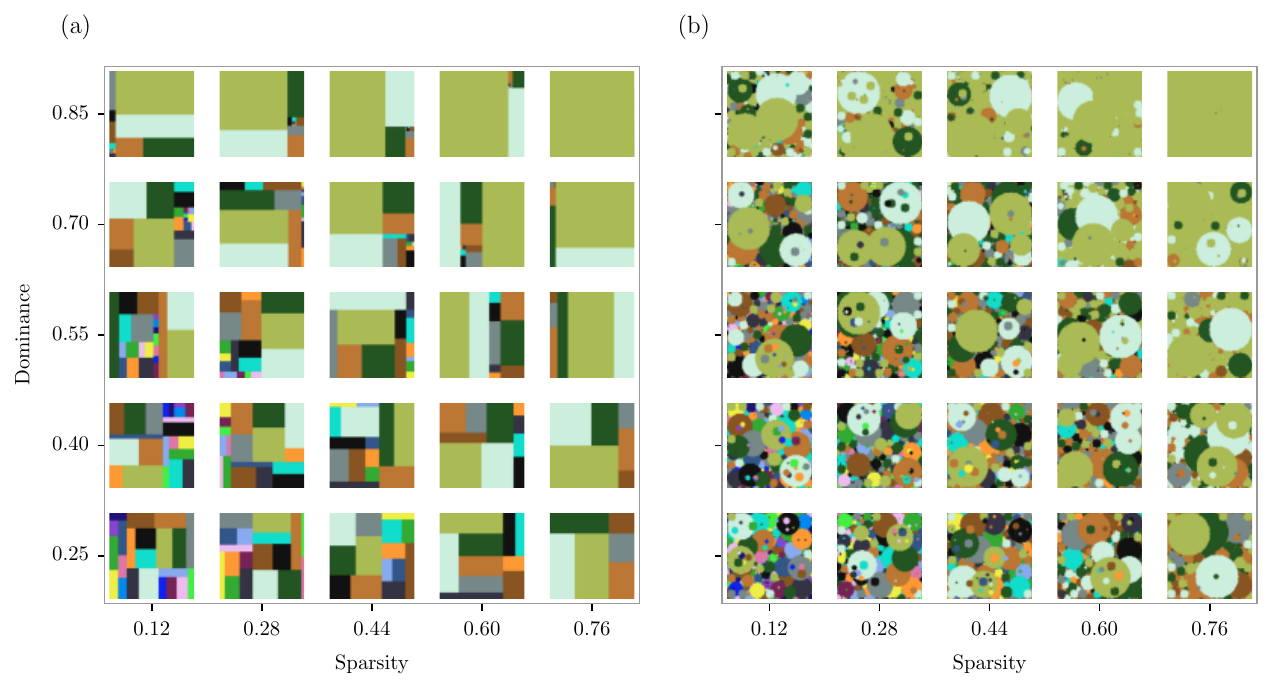}
    \caption{Stimuli generated from the palette of C\'ezanne's \emph{Mont Sainte-Victoire} (1885), as in Fig.~\ref{fig:survey_imagesP1}.}
    \label{fig:survey_imagesP2}
\end{figure}

\begin{figure}[H]
    \centering
    \includegraphics[width=\linewidth]{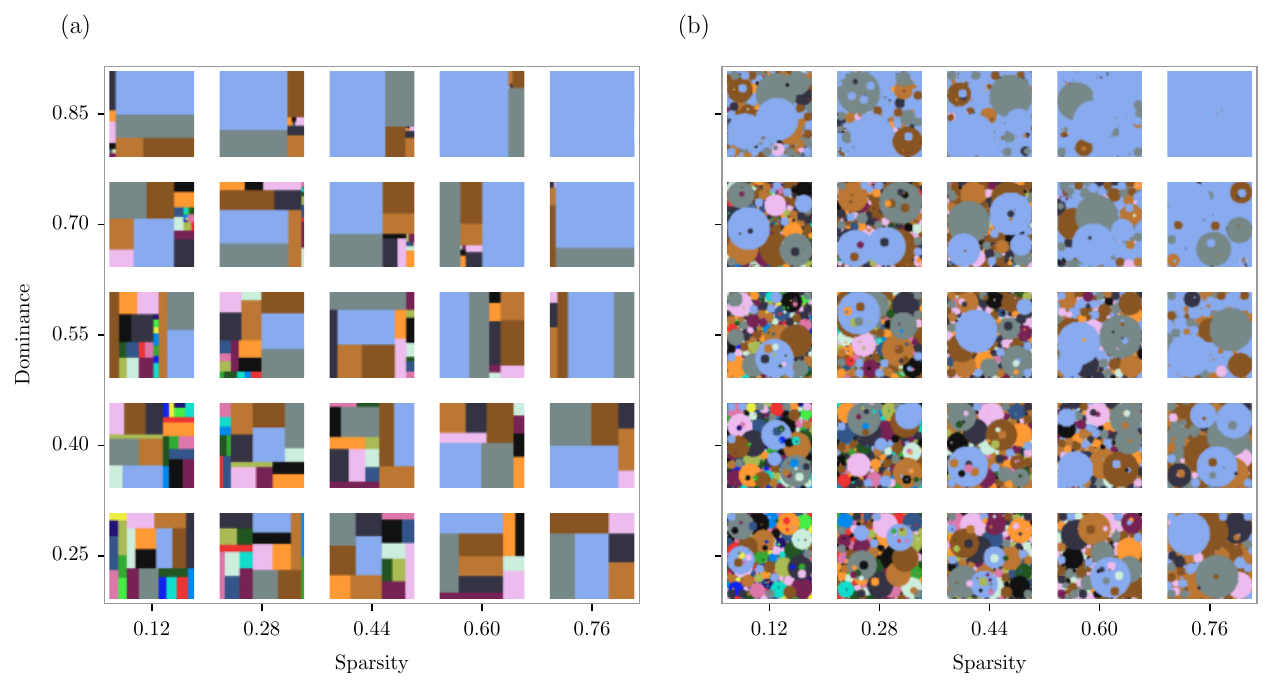}
    \caption{Stimuli generated from the palette of Raphael's \emph{The Triumph of Galatea} (1512), as in Fig.~\ref{fig:survey_imagesP1}.}
    \label{fig:survey_imagesP3}
\end{figure}

\begin{figure}[H]
    \centering
    \includegraphics[width=\linewidth]{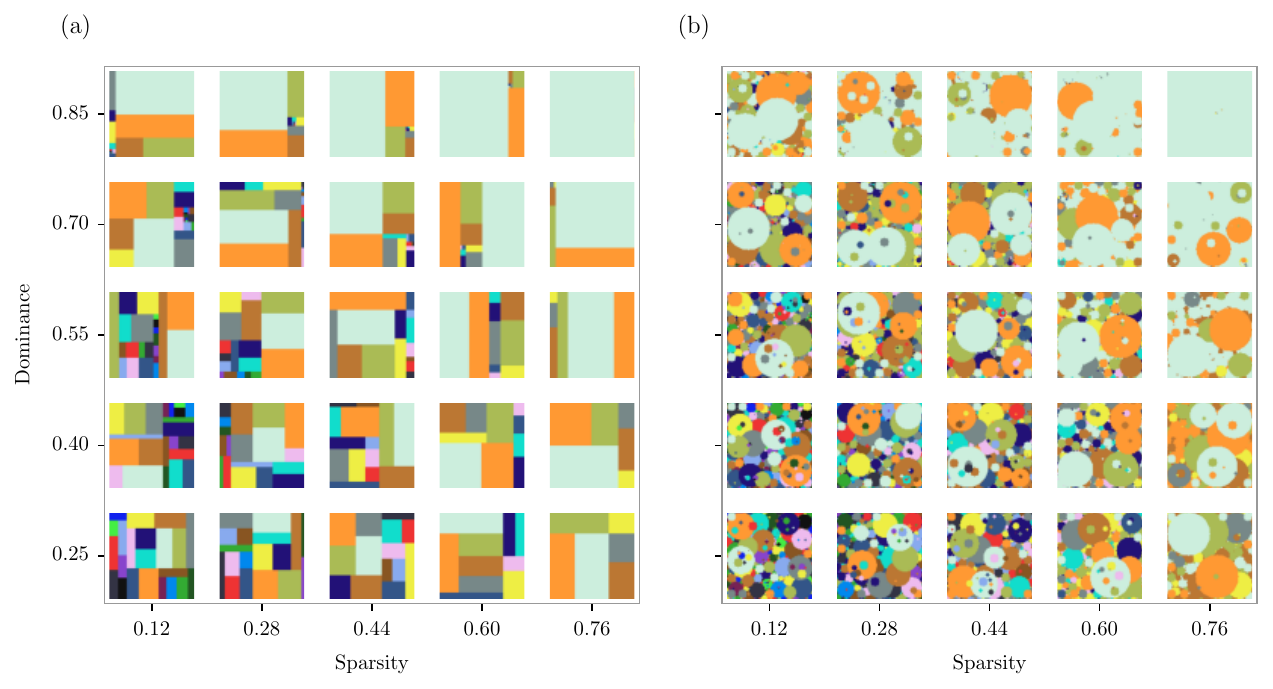}
    \caption{Stimuli generated from the palette of Rothko's \emph{Multiform} (1948), as in Fig.~\ref{fig:survey_imagesP1}.}
    \label{fig:survey_imagesP4}
\end{figure}

\begin{figure}[H]
    \centering
    \includegraphics[width=\linewidth]{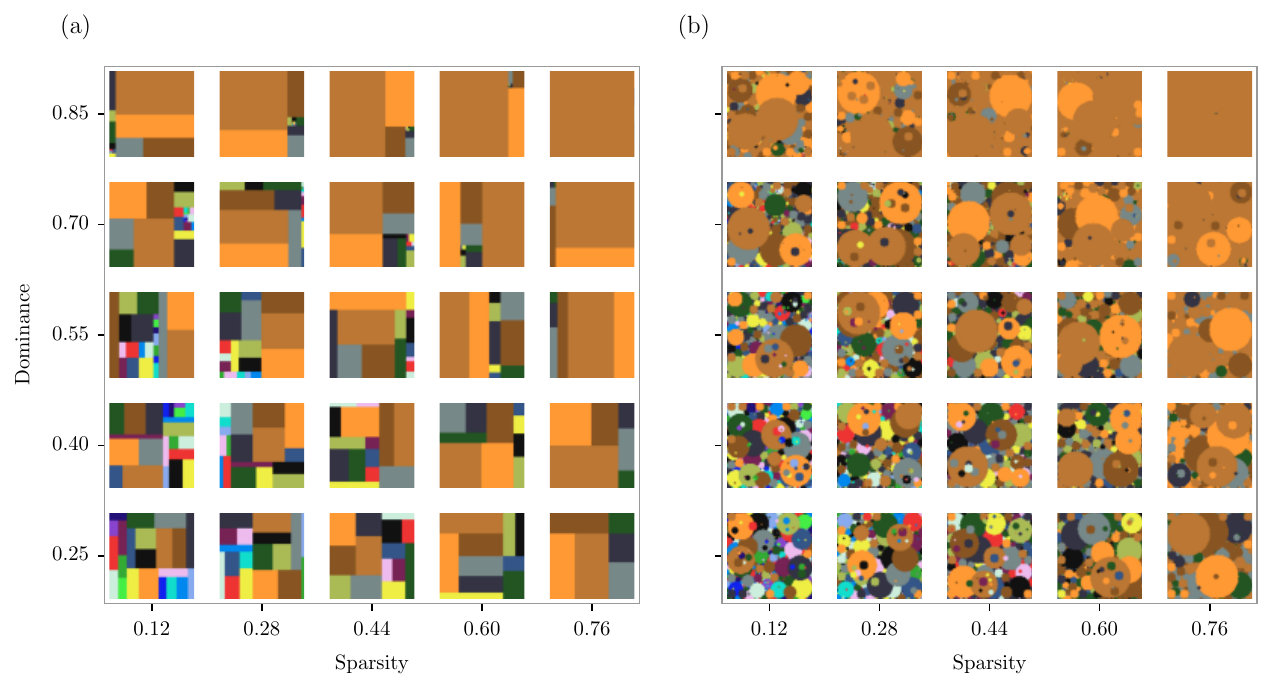}
    \caption{Stimuli generated from the palette of Turner's \emph{The Burning of the Houses of Parliament} (1834), as in Fig.~\ref{fig:survey_imagesP1}.}
    \label{fig:survey_imagesP5}
\end{figure}

\paragraph{Rectangular patches.} In the first survey, each image is a partition of the canvas into non-overlapping rectangular patches, one per active color, with areas proportional to the prescribed proportions (panel (a) of Figs.~\ref{fig:survey_imagesP1}--\ref{fig:survey_imagesP5}). The spatial layout was drawn at random and then held fixed, so that images generated from different paintings at the same $(s,d)$ share an identical geometry and differ only in their colors. Any difference in response between them is therefore attributable to color alone.\\

\paragraph{Circular patches.} In the second survey, the pixel count assigned to each active color is split among several circles, each defined by a radius and a center drawn at random on the canvas, with overlaps allowed (panel (b) of Figs.~\ref{fig:survey_imagesP1}--\ref{fig:survey_imagesP5}). Radii are bounded between 5 and 25 pixels. The lower bound keeps patches large enough to be perceived as color rather than noise, while the upper bound, a diameter equal to half the image side, prevents a single patch from dominating the composition. The number of circles per color is capped at $n_{\max}$. For $n_{\max}\lesssim5$ the circles are uniformly large and the lower radius bound is rarely reached, while for $n_{\max}\gtrsim10$ the count is fragmented into small patches that rarely approach the upper bound. The count is then divided among the circles by a symmetric Dirichlet distribution of concentration parameter $\alpha$, with $\alpha<1$ favoring uneven allocations and hence heterogeneous circle sizes. We use $(n_{\max},\alpha)=(6,0.1)$, which provides sufficient heterogeneity while keeping the images visually coherent. Since overlaps change the visible pixel counts, a final correction step adjusts the patches until the measured $(s,d)$ matches its target as closely as possible. The correction radius is capped at 5 pixels, in line with the minimum circle radius, so that corrections act only through minimal local adjustments.\\

\paragraph{Survey structure.} Each question displays five images, obtained by holding one of the two coordinates fixed at one of its five grid values and letting the other run over its full range, and asks the participant to select the one with the most harmonious color balance. This yields ten questions per painting, five at fixed sparsity and five at fixed dominance, and hence a pool of fifty distinct questions. Each participant answers ten of them, drawn at random under three constraints. Exactly two questions come from each painting, each of the ten fixed-value conditions appears exactly once, and no two consecutive questions fix adjacent grid values, to avoid serial anchoring. The order of the five images within a question is randomized independently for every participant. To check that the spatial arrangement does not itself drive the choice, an eleventh question was appended, showing five images that share the same pair $(s,d)$ and differ only in the placement of their patches. For each participant it was drawn from a pool of four, generated from the palette of \emph{The Starry Night} at the pairs $(0.12,0.25)$, $(0.12,0.85)$, $(0.44,0.55)$ and $(0.76,0.25)$. The question was presented exactly like the others, with no indication that it served as a control, and no layout was selected appreciably more often than the rest, confirming that the responses are driven by the color proportions rather than by their placement.

\clearpage

\end{document}